%% file: main.tex
\documentclass{article}

\usepackage{natbib}
\usepackage{amsmath}
\usepackage{amssymb}
\usepackage{graphicx}
\usepackage[utf8]{inputenc}
\usepackage[T1]{fontenc}
\usepackage{url}
\usepackage{booktabs}
\usepackage{amsfonts}
\usepackage{nicefrac}
\usepackage{microtype}
\usepackage{wrapfig}
\usepackage{hyperref}

\usepackage{subcaption}
\usepackage{arxiv}

\usepackage{titlesec}
\titlespacing\section{0pt}{0pt plus 2pt minus 2pt}{0pt plus 2pt minus 2pt}
\titlespacing\subsection{0pt}{3pt plus 4pt minus 2pt}{0pt plus 2pt minus 2pt}
\titlespacing\subsubsection{0pt}{3pt plus 4pt minus 2pt}{0pt plus 2pt minus 2pt}

\title{ProtoTransformer: A Meta-Learning Approach to Providing Student Feedback}

\author{
    Mike Wu$^{1}$, Noah Goodman$^{1,2}$, Chris Piech$^{1}$, Chelsea Finn$^{1}$\\
    Department of Computer Science$^{1}$ and Psychology$^{2}$\\
    Stanford University\\
    Stanford, CA 94305\\
    \texttt{\{wumike, ngoodman, cpiech, cbfinn\}@stanford.edu}
}

\begin{document}

\maketitle

\input{sections/abstract}
\input{sections/introduction}
\input{sections/background}
\input{sections/realworld}
\input{sections/methods}

\input{sections/nlp}
\input{sections/experiments}

\input{sections/deployment}
\input{sections/related}
\input{sections/conclusion}

\bibliography{references}
\bibliographystyle{plainnat}

\input{sections/appendix}

\end{document}

%% file: sections/abstract.tex
\begin{abstract}
     High-quality computer science education is limited by the difficulty of providing instructor feedback to students at scale. While this feedback could in principle be automated, supervised approaches to predicting the correct feedback are bottlenecked by the intractability of annotating large quantities of student code. In this paper, we instead frame the problem of providing feedback as few-shot classification, where a meta-learner adapts to give feedback to student code on a new programming question from just a few examples annotated by instructors. Because data for meta-training is limited, we propose a number of amendments to the typical few-shot learning framework, including task augmentation to create synthetic tasks, and additional side information to build stronger priors about each task. These additions are combined with a transformer architecture to embed discrete sequences (e.g. code) to a prototypical representation of a feedback class label. On a suite of few-shot natural language processing tasks, we  match or outperform state-of-the-art performance.  Then, on a collection of student solutions to exam questions from an introductory university course, we show that our approach reaches an average precision of 88\% on unseen questions, surpassing the 82\% precision of teaching assistants. Our approach was successfully deployed to deliver feedback to 16,000 student exam-solutions in a programming course offered by a tier 1 university. This is, to the best of our knowledge, the first successful deployment of a machine learning based feedback to open-ended student code.
\end{abstract}

%% file: sections/introduction.tex
\section{Introduction}

High quality education at scale is a long-standing unsolved challenge that is only becoming more important. The price of education per student is growing faster than economy-wide costs \citep{bowen2012cost}, bounding the amount of resources available to support individual student learning. However, the demand for higher education is steadily increasing \citep{kumar2015supply}.
This is especially urgent for computer science education as workers in traditional jobs are being displaced and require re-skilling.
Massive open online courses, or MOOCs, have responded to this challenge by democratizing access to high quality content.
But, content is only a part of the solution. MOOCs notoriously suffer from high rates of student disengagement and dropout \citep{kizilcec2013deconstructing} -- only a small fraction of students who start the courses finish.
This is in part because MOOCs  ignore another important ingredient for learning: \textit{feedback}.
Focusing on computer science education, this paper studies the problem of providing feedback on student code, a challenging process that is critical to learning but requires expertise and is difficult to scale due to the vast diversity of questions and student responses.
Existing autonomous methods such as unit tests can help recognize a correct solution; however they are not especially helpful at providing \emph{feedback} which could guide a student to better understanding. As such, well resourced courses typically spend a huge amount of human effort manually providing feedback to student code.

With datasets of student work becoming more commonplace \citep{settles2018data,husain2019codesearchnet,riiid2020}, there is an opportunity to use machine learning to provide feedback at scale.
But, naively treating this as a standard supervised learning problem  faces challenges with overfitting to small datasets and trouble generalizing to new students and questions \citep{wu2019zero, basu2013powergrading,yan_pyramid,wang2017learning,liu2019automated,hu2019reliable}.
In order for a method to be applicable in the real world, it must use expert annotations more efficiently.

\begin{wrapfigure}{l}{0.5\linewidth}
\centering
\includegraphics[width=\linewidth]{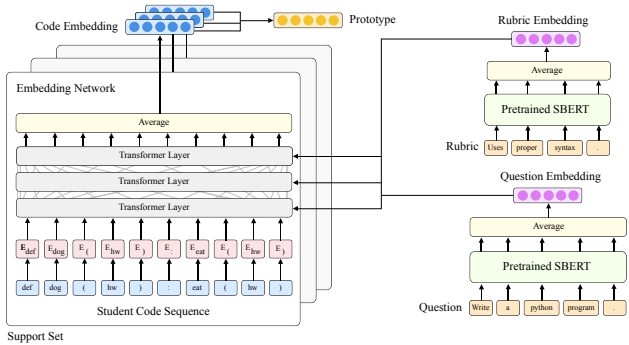}
\caption{{\small\textbf{ProtoTransformer Networks for Few-Shot Classification of Student Code:}
Given a programming question, the ProtoTransformer Network is trained to predict feedback for student code using only a small set of annotated examples.
Feedback categories are specified according to a rubric e.g. ``Incorrect syntax'' or ``Missing variable''.
Question and rubric descriptions are embedded with pretrained SBERT.
Student code is then encoded through stacked transformer layers conditioned on question and rubric embeddings.
A ``prototype'' is the average code embedding  for each class label. New examples are embedded and compared to each prototype.}}
\label{fig:model}
\end{wrapfigure}

We therefore formulate feedback to students as a few-shot classification problem: given a handful of annotated examples of student solutions to a new question, a model must quickly adapt to provide feedback to new student responses.
Meta-learning approaches have demonstrated promising results on such problems by learning how to quickly learn a classifier from a few examples. However, prior meta-learning methods have largely focused on few-shot visual recognition tasks constructed by repurposing existing datasets  \citep{lake2015human,vinyals2016matching,snell2017prototypical,finn2017model,yu2018one}.
In contrast, the feedback problem presents a number of new challenges that arise in real-world applications including imbalanced data distributions, limitations on meta-learning data size, and shifts in the distribution of tasks. Together, these challenges make it nontrivial to apply existing meta-learning algorithms.

The main contribution of this work is a meta-learning framework for few-shot classification of sequence data, including programming code, that aims to mitigate these challenges.
We propose an augmentation technique for code data that can generate additional tasks for meta-learning.
To handle the ambiguity of learning from only a few examples and better guide the few-shot learner, we incorporate side information.
To address the exponential number of unique variable and function names in code sequences, we propose to use byte-pair encoding or program obfuscation.
All of these components are integrated with an architecture that uses transformers \citep{vaswani2017attention} to compute prototypical embeddings of a class label.

To benchmark our model in existing meta-learning contexts, we first study a suite of few-shot text classification tasks, where we find our approach to rival, if not outperform, several state-of-the-art meta-learning NLP algorithms.
Then, on a dataset of student solutions to university exams, our approach achieves an average precision (AP) of 88\% when providing feedback to unseen problems, compared to the 82\% AP from teaching assistants and 65\% AP from standard supervised approaches.
That is, our model rivals the average human instructor, opening opportunities for real-world impact in the classroom.
We also include a detailed ablation study to measure the impact of individual components of our approach, and exhaustively compare to alternative choices.
Finally, we deployed our approach to provide feedback to 16,000 student solutions for a diagnostic exam in a large online programming course, demonstrating the immediate impact possible.

%% file: sections/background.tex
\section{Preliminaries}
\label{sec:background}
We review the basics of few-shot learning and introduce the feedback challenge.


\textbf{Few-Shot Learning Setup}$\quad$
Suppose we are given a collection of \textit{tasks} containing examples for each of $N$ classes.
Each task is divided into a \textit{support set} of $K$ examples per class, $\mathcal{S} = \{(x_1, y_1), (x_2, y_2), \ldots, (x_{K\times N}, y_{K\times N})\}$, and a \textit{query set} of $Q$ examples from the same $N$ classes, $\mathcal{Q} = \{(x^*_1, y^*_1), (x^*_2, y^*_2), \ldots, (x^*_{Q\times N}, y^*_{Q\times N})\}$.
For simplicitly, we assume $N$ is the same across all tasks.
Every $x_i$ represents a data example (e.g. image, text, or vector), and every $y_i \in \{1, 2, \ldots, N\}$ is a class label. The challenge of few-shot learning is to fit a model that uses the support set $\mathcal{S}$ to predict labels on the query set $\mathcal{Q}$ for every task. Tasks are divided into a meta-training set and a meta-test set. The model parameters are chosen based on performance on the meta-training set. In evaluation, the model is judged by performance on the query set of each task in the meta-test set.

\textbf{Prototypical Networks}$\quad$ Prototypical Networks \citep{snell2017prototypical} learn an embedding function $f_\theta$ that maps every example $x_i$ to a vector in $\mathbb{R}^d$. In practice, $f_\theta$ is a deep neural network where $\theta$ represents trainable parameters. This embedding is used to map the support set $\mathcal{S}$ to define a \textit{prototype embedding} for each class (usually the average embedding over $K$ shots). The goal is for the embedding of query examples to be closer to the prototype of their correct class than alternative classes.

Precisely, the prototype for the $c$-th class is $p_c = \frac{1}{K} \sum_{x_i \in \mathcal{S}_c} f_\theta(x_i)$ where $\mathcal{S}_c$ is the subset of the support set $\mathcal{S}$ with examples labeled with class $c$. Then, the prototypical network objective is to minimize a softmax over distances of each query example $(x^*, y^*) \in \mathcal{Q}$ to each prototype:
\begin{equation}
    \mathcal{L}(x^*, y^*) = -\log \frac{\exp\{-\textup{dist}(f_\theta(x^*), p_{y^*}) / \tau\}}{\sum_{c=1}^N\exp\{-\textup{dist}(f_\theta(x^*), p_c) / \tau\}}
\label{eqn:prototype}
\end{equation}
where $\textup{dist}(\cdot, \cdot)$ represents L$_2$ distance.
The parameter $\tau$ can be either a constant or learned \citep{oreshkin2018tadam}.
Eq.~\ref{eqn:prototype}, averaged over all query examples, is minimized with stochastic gradient descent.
At meta-test time, the support set $\mathcal{S}$ is again embedded to construct prototypes, following which a label is predicted for a query example $x^*$ by finding the closest prototype to its embedding, $f_\theta(x^*)$.

\begin{figure}[h!]
\centering
\includegraphics[width=0.75\linewidth]{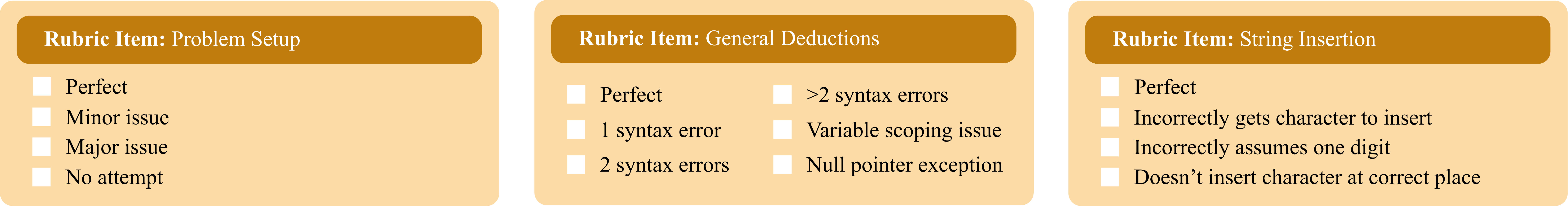}
\caption{\textbf{Feedback rubrics:} a rubric contains multiple items, each of which has a text description and several options. Rubric options are varied in granularity and number.}
\label{fig:rubric}
\end{figure}

\textbf{The Feedback Challenge}$\quad$ To introduce the feedback challenge, suppose we have access to student solutions to a set of programming exercises. Each student response is annotated with a \textit{rubric}, which contains several \textit{rubric items}, each describing a student misconception. For example, if the problem required the student to use a ``for loop'', rubric items may explore misunderstandings with the iterator, the termination condition, or general syntax. Each rubric item contains a text description and several \textit{rubric options} that an annotator may pick, varying from a ``perfect solution'' to multiple misconception types. More than one option may be chosen for a single rubric item. See Figure~\ref{fig:rubric} for example rubrics. Notably, every programming question has its own rubric with mostly unique items and options. The goal of the feedback challenge is to predict the correct assignments for every rubric item from student code with as little required annotation as possible.

The hypothesis of this paper is tackle the feedback challenge through few-shot learning of student prototypes.
Our motivation for learning prototypes comes from a  close relationship to applications of item response theory in examination \citep{edgeworth1888statistics}, a popular framework that measures student ability and question difficulty by averaging over responses (i.e. a simple prototype).
This is contrast with most previous attempts at feedback \citep{wu2019zero,malik2019generative}, which leverage mostly hand-crafted models.
Despite their success on block-based programs, scaling feedback in education is still an unsolved real-world problem.
We see an important contribution of ours as carefully formulating student feedback as a machine learning problem, and more specifically a few-shot meta-learning problem.
With a performant few-shot feedback system, instructors could cheaply provide feedback for $K$ students, allowing the automated system to provide feedback for the rest.

%% file: sections/realworld.tex
Although we have explained the crux of the problem, there are multiple challenges that persist in real-world applications that make the feedback challenge so difficult.

\textbf{Limitations on Task Annotation}$\quad$
Unlike labeling object classes in an image, labeling domain-specific data requires field expertise and intensive labor.
To make matters worse, meta-learning algorithms require not just one dataset, but a set of datasets, one for each task.
Whereas benchmark meta-learning datasets, e.g. Omniglot or miniImagenet, leverage thousands of tasks, limited annotation in real world settings could restrict the number of training tasks to be a few hundred at best.
In computer science education, grading programming code requires the ``annotator'' to infer the core misconception made by the student.
This complicated inference task is as least as difficult as debugging someone else's code, as the reader must understand the student's thought process and trace errors back to the source.
Unsurprisingly, this is incredibly hard to scale.
Wu et. al. \citep{wu2019zero} found that annotating 800 Blockly programs took university teaching assistants 25 human-hours.


\textbf{Long-tailed Data Distributions}$\quad$
Real world datasets face a challenging problem of data imbalance.
Several classes might be much more rare than others, such as uncommon coding mistakes students make on a class exercise.
Strong class imbalance limits the size of the support set available to train a meta-learning model.
Rare classes also pose a more difficult generalization challenge as examples in the long-tail of the data distribution are very diverse.
In an average classroom, students have an exponential number of ways to approach the same problem. For a coding assignment, the distribution of student code has been shown to be approximately Zipfian \citep{wu2019zero}.
Despite this, instructors deeply care about feedback in the  ``long-tail'', as it contains struggling students.


\textbf{Side Information}$\quad$
In real-world contexts, we often have access to weak auxiliary information about a task, such as a description, a webpage, or metadata.
However, many few-shot learning methods are deprived of this information, operating only on the support and query examples.
Careful consideration of such ``side information'' could reduce ambiguity of individual tasks that arises from limited data.
In education, graded student work can be accompanied by a description of the question, the rubric for grading, and even metadata about the student, grader, or course.


%% file: sections/methods.tex
\section{ProtoTransformer Networks}
\label{sec:methods}
In this section, we introduce a few-shot learning framework with several components designed to help address the above challenges. We start by introducing the ProtoTransformer architecture, which can process sequence data points and readily incorporate side information to solve a few-shot task.
Following, we introduce two components designed to
address limited and long-tailed distributions of programming code: task augmentation and variable name encodings.

As in Section~\ref{sec:background}, we are given a support set $\mathcal{S}$ and a query set $\mathcal{Q}$ for every task. Additionally, we now assume that every example $x_i$ is a sequence of discrete tokens i.e. $x_i = (x_{i,1}, x_{i,2}, \ldots, x_{i,T})$
where each token $x_{i,t} \in [1, V]$ is one of $V$ words in a vocabulary and $T$ is a maximum length.
In the ProtoTransformer, we parameterize the embedding network $f_\theta$ as stacked transformer encoder layers as presented in RoBERTa \citep{liu2019roberta} (see Figure~\ref{fig:model}).
The RoBERTa architecture maps a sequence of tokens to a sequence of embeddings, which we aggregate into a single embedding by averaging over non-padding tokens.
Given this average embedding, we compute prototypes and optimize Eq.~\ref{eqn:prototype}.
Even for the prototypical objective, we found it important to use Adam with $\beta_1$ = 0.9, $\beta_2$ = 0.999, $\epsilon$ = 1e-6, and warm up the learning to 1e-4, followed by linear decay, as in \citep{liu2019roberta}.

\subsection{Conditioning on Side Information}

For a given task, we may have some side information $z$, along with the support and query sets.
We assume side information $z$ is known apriori as a fixed discrete sequence of tokens, $z = (z_1, z_2, .., z_T)$.
Further, we assume access to a second embedding function $g_\phi$ that maps $z$ to a vector.
For instance, if $z$ is a English description, we may choose $g_\phi$ to be a pretrained sentence model, such as SBERT \citep{reimers2019sentence}.

\begin{figure}[h]
\centering
\includegraphics[width=0.75\linewidth]{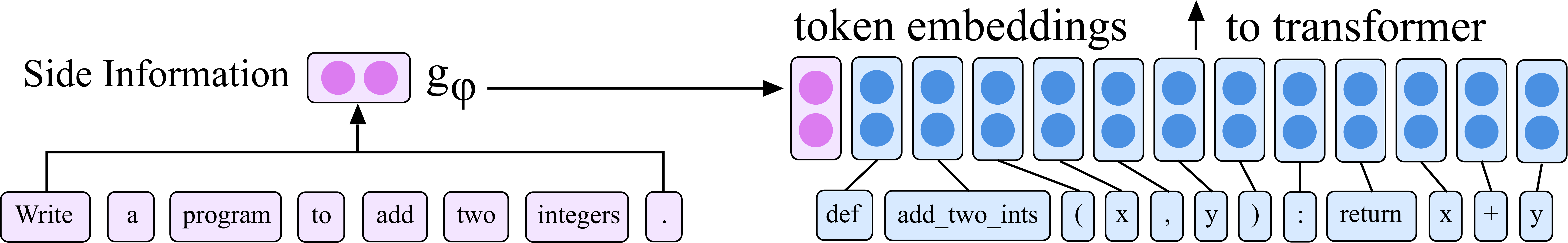}
\caption{\textbf{Side Information as a ``Task Token'':} Side information (pink) is embedded by a network $g_\phi$ and prepended to token embeddings (blue). The combined sequence is fed to the transformer.}
\label{fig:tam2}
\end{figure}

A question remains on how to incorporate side information vectors $g_\phi(z)$ into the embedding function $f_\theta$ so that the model can fuse information from the support set and the side information together.
We propose a very simple approach:
first, the side information embedding $g_\phi(z)$ is transformed to be the same dimensionality as the input embedding in $f_\theta$ (e.g. 768 for RoBERTa) using a linear layer.
Then, we prepend this resulting vector to each input sequence in the support set and query set.
That is, we treat $g_\phi(z)$ as the embedding of a special ``task token''.
This new input sequence is provided to the stacked transformers, in which the attention layers will mix the side information into the final token embeddings.
See Figure~\ref{fig:tam2} for a illustration.
In our experiments, we compare performance with and without side information, finding a 2-4 point increase by including it.

\subsection{ProtoTransformers for Code}
\label{sec:code}

When applying our approach to few-shot classification of programming code, we introduce technical innovations specific to code but not to any single programming language nor application domain.

\textbf{Standardizing Code Input } While it is common to use abstract syntax trees \citep{wu2019zero,jain2020contrastive,alon2018code2seq}, we opt for a simpler standardization that is more faithful to the original program string: we use the built-in parser to deconstruct the string into tokens (e.g. \texttt{pythonlang.tokenize} for Python). The resulting output tags segments of code with semantic labels, such as comments, names, numerics, symbols, etc. For every program, we remove all inline comments, convert all variable and function names from camel to snake case, replace newlines with a unique symbol (\texttt{<newline>}), and explicitly demarcate entering and exiting a new scope with unique symbols (e.g. \texttt{<scope>} and \texttt{</scope>}). Note that programs with syntax errors can still be tokenized in this manner.

\textbf{Self-Supervised Tasks for Code } We explore two approaches of constructing synthetic tasks from unlabeled programming code.
First, inspired by subset masked language modeling tasks, or SMLMT \citep{bansal2020self}, we similarly wish to utilize an ``un-masking'' task.
As SMLMT was designed for natural language, it has a subtle shortcoming when applied to programming code: as a significant portion of tokens in a program are used for function and variable names, randomly masking tokens through SMLMT will often mask these naming tokens.
Unfortunately, naming tokens have little to do with program syntax nor structure, and can be arbitrary.
That is, multiple choices can be equally as good despite only one being the true token.
For example, \texttt{sum}, \texttt{z}, or \texttt{total} are all good choices for un-masking \texttt{<mask> = x + y}.
Therefore, we propose a separate ``cloze'' task to work around this issue: we do \textit{not} mask function or variable names to focus on program semantics.
We use the parser to identify naming tokens. Then, we randomly choose $N$ tokens (e.g. \texttt{def} or \texttt{range} in Python) and $K + Q$ examples for each class that contain the chosen tokens --- together, these compose a $N$-way classification problem.
For each program, all instances of the chosen token are replaced with \texttt{<mask>}.
In Section~\ref{sec:experiments}, we compare our cloze task to the SMLMT task and find 3-4 points of improvement.

We propose a second synthetic task by predicting compilation errors.
In Python, all programs can be compiled with the \texttt{pythonlang.tokenize} function.
Given a set of possible compilation outcomes e.g. \texttt{TypeError}, \texttt{SyntaxError}, or \texttt{IndentationError}, we construct a task by randomly choosing $N$ outcomes, and picking $K+Q$ programs that compile to each of the classes.
Successfully predicting compilation error requires the model to understand program logic.
Alternatively, if one had access to unit tests, they could be used in addition to compilation tasks.


\textbf{Variable and Function Names } A subtle difficulty of working with programming code is handling variable and function names, which are responsible for more than 99\% of the unique vocabulary.
Naively training the ProtoTransformer on programs with their original variable and function names is challenging.
since unique names in unseen programs make it very hard for the embedding function to generalize. We explore two approaches to ``normalize'' naming.

The first approach is to take advantage of byte-pair encodings (BPE). Given we have converted all variable and function names to snake case, most naming is comprised of english words conjoined by underscores (e.g. \texttt{my\_variable}).
As such, byte-pairs of programming names resemble standard English.
To this end, we can leverage the RoBERTa tokenizer \citep{liu2019roberta}.


A second approach is to obfuscate variable and function names. To do this, we allot the model $N_v$ variable name tokens (e.g. \texttt{<var:1>}, $\ldots$, \texttt{<var:$N_v$>}) and $N_f$ function name tokens (e.g. \texttt{<func:1>}, $\ldots$, \texttt{<func:$N_f$>}).
In meta-training, every support or query example is transformed to replace all names with randomly chosen name tokens, in a consistent manner. The randomness is important so as not to memorize relationships between specific name tokens and positions in code. In meta-test, we sequentially replace variable and function names from left-to-right. The motivation is to provide the model the expressivity to represent variables and functions but minimize the vocabulary size.

\textbf{Pretraining on Unlabeled Data } Given a large corpus of unlabeled programming code, we learn an unsupervised representation that captures semantic information on code, which could be useful for solving few-shot classification tasks:
since CodeBERT \citep{feng2020codebert} also use stacked transformer layers, we can initialize the ProtoTransformer Network with its pretrained weights, and finetune only the top few layers with Eq.~\ref{eqn:prototype}.
Since the unlabeled corpus is order of magnitudes larger than the annotated one, we found initializing weights to  improve meta-learning capability by up to 30 points.

%% file: sections/nlp.tex
\section{Few-Shot Natural Language Processing Experiments}

Before we focus on educational feedback, we recognize that education is a relatively new application for machine learning with limited public data due to student privacy.
At the same time, our approach is not specific to code nor education. We find side information and task augmentation benefit meta-learning more generally.
To compare our approach in a more familiar context, we study few-shot text classification problems in natural language processing, and find encouraging results.
\begin{table}[h]
    \tiny
    \centering
    \begin{subtable}[h]{0.50\textwidth}
        \centering
        \begin{tabular}{lcccc}
        \toprule
        $N$ & \multicolumn{2}{c}{2-way} & \multicolumn{2}{c}{3-way} \\
        \midrule
        Model & 1-shot & 2-shot & 1-shot & 2-shot \\
        \midrule
        All (ours) & \textbf{85.5} $\pm$ 15 & \textbf{87.0} $\pm$ 17 & \textbf{77.9} $\pm$ 11 & \textbf{82.3} $\pm$ 07 \\
        No Pretrain & 55.5 $\pm$ 14 & 62.5 $\pm$ 11 & 57.9 $\pm$ 15 & 57.6 $\pm$ 12 \\
        No Side & 83.5 $\pm$ 16 & 85.5 $\pm$ 20 & 73.3 $\pm$ 08 & 79.3 $\pm$ 09 \\
        No SMLMT & 84.0 $\pm$ 20 & 85.5 $\pm$ 21 & 72.3 $\pm$ 11 & 76.9 $\pm$ 12 \\
        LSTM & 52.5 $\pm$ 14 & 52.5 $\pm$ 09 & 47.9 $\pm$ 09 & 47.3 $\pm$ 10 \\
        Matching & -- & 84.5 $\pm$ 17 & -- & 81.6 $\pm$ 16 \\
        Supervised & 53.0 $\pm$ 23 & 57.0 $\pm$ 12 & 35.8 $\pm$ 06 & 39.9 $\pm$ 08 \\
        \bottomrule
        \end{tabular}
    \end{subtable}
    \begin{subtable}[h]{0.49\textwidth}
        \centering
        \centering
        \begin{tabular}{ll}
        \toprule
        \textbf{Model} & \textbf{Meta-Test Acc.} \\
        \hline
        Matching Network \citep{vinyals2016matching} & 65.73 \\
        Prototypical Network \citep{snell2017prototypical} & 68.17 \\
        Graph Network \citep{garcia2017few} & 82.61 \\
        Relation Network \citep{sung2018learning} & 83.07 \\
        SNAIL \citep{mishra2017simple} & 82.57 \\
        ROBUSTTC-FSL \citep{yu2018diverse} & 83.12 \\
        Induction Network \citep{geng2019induction} & \textbf{85.63} \\
        ProtoTransformer (Ours) & \textbf{85.07} ($\pm$ 2.4) \\
        \hline
        ProtoTransformer (No Pretrain) & 75.86 ($\pm$ 2.5) \\
        ProtoTransformer (No Side Info) & 83.67 ($\pm$ 2.0)\\
        ProtoTransformer (No Cloze) & 84.82 ($\pm$ 2.2)\\
        ProtoTransformer (LSTM) & 77.67 ($\pm$ 2.1) \\
        \bottomrule
        \end{tabular}
    \end{subtable}
    \caption{{\small \textbf{Few-shot topic classification (left):} ProtoTransformer performance on few-shot topic classification using the 20-newsgroups dataset, varying the number of shots and classes. \textbf{Few-shot sentiment classification (right):} we matches state-of-the-art performance on 5-shot ARSC multi-domain sentiment classification.}}
    \label{table:nlp}
    \vspace{-2em}
\end{table}

\textbf{Experiment Setup}$\quad$ First, we re-purpose the 20-newsgroups dataset, which contains posts on 20 topics (such as sports, politics, or religion) for few-shot topic classification. To build a task, we randomly choose $N$ topics and pick $K+Q$ examples per topic (We vary $N$ and $K$ but fix $Q$=10). We reserve 5 topics for meta-testing randomly.
To incorporate side information, we provide the model the SBERT embedding of the $N$ topic names.
Second, we study few-shot sentiment analysis using the Amazon Reviews (ARSC) dataset \citep{blitzer2007biographies}, replicating \cite{geng2019induction} to construct 69 five-shot binary classification tasks. Here, we leverage the product  name (e.g. books, DVDs, or electronics) as side information. Finally, we evaluate our model with a suite of five text classification datasets proposed by \cite{bao2019few}. Unless ablated, we encode text with BPE, use a pretrained RoBERTa, and add 10\% SMLMT tasks.

\begin{table}[h]
    \scriptsize
    \centering
    \begin{tabular}{llllll}
    \toprule
    \textbf{Model} & \textbf{20News} & \textbf{Amazon} & \textbf{HuffPost} & \textbf{Reuters} & \textbf{FewRel}\\
    \hline
    MAML \citep{finn2017model} & 43.9 & 47.2 & 49.6 & 62.5 & 57.8 \\
    Proto. Net \citep{snell2017prototypical} & 45.4 & 51.9 & 41.6 & 68.1 & 46.5 \\
    Ridge \citep{bertinetto2018meta} & 57.2 & 72.7 & 54.8 & 90.0 & 72.2 \\
    Signatures \citep{bao2019few} & \textbf{68.3} & 81.1 & 63.5 & 96.0 & \textbf{83.5} \\
    ProtoTransformer & 61.3 ($\pm$0.4) & \textbf{85.2} ($\pm$0.3) & \textbf{65.0} ($\pm$0.3) & \textbf{97.7} ($\pm$0.1) & 75.9 ($\pm$0.3) \\
    \bottomrule
    \end{tabular}
    \caption{{\small Suite of \textbf{5-way 5-shot text classification} tasks. The ProtoTransformer Network closely matches state-of-the-art algorithms, outperforming them in 3 out of 5 datasets.}}
    \label{table:nlp:3}
    \vspace{-0.4cm}
\end{table}
Table~\ref{table:nlp} (left) reports the meta-learning approach to outperform a supervised baseline by 32 points absolute. Table~\ref{table:nlp} (left) also reports several ablations isolating the impact of different components of our approach. We find large improvements of 30 points by utilizing transformers and pretraining as well as 2-4 points of improvement through side information and augmented tasks.
Table~\ref{table:nlp} (right) and Table~\ref{table:nlp:3} show that the ProtoTransformer rivals state-of-the-art methods. Induction Networks and ProtoTransformers are comparable on ARSC, surpassing SNAIL, graph, and relation networks by 3\% absolute and standard prototypical networks by 17\% absolute. On \cite{bao2019few}'s suite of datasets, we  outperform Distributional Signatures \citep{bao2019few} on 3 out of 5 datasets, also surpassing MAML and Prototypical Networks by 20-30\% absolute. In light of these results, we view the ProtoTransformer as a modernization of the Prototypical Network --- with straightforward extensions, the prototype approach remains competitive with more sophisticated meta-learning algorithms for natural language.

%% file: sections/experiments.tex
\section{Scaling Feedback for University Exams}
\label{sec:experiments}
In the coming experiments, we are interesting in the following questions: \textbf{(1)} How does feedback from few-shot learning with the ProtoTransformer Network compare to human feedback?
\textbf{(2)} In the low data meta-learning setting, do  constructing synthetic tasks, adding side information, and careful data preprocessing improve performance? Which ones are more important?


\begin{wraptable}{l}{.5\linewidth}
    \vspace{-0.2cm}
    \scriptsize
    \centering
    \begin{tabular}{lcccc}
    \toprule
    &  \multicolumn{4}{c}{Held-out rubric} \\
    Model & AP & P@50 & P@75 & ROC-AUC \\
    \midrule
    ProtoTransformer & \textbf{84.2} & \textbf{85.2} & \textbf{74.2} & \textbf{82.9} \\
    & ($\pm$1.7) & ($\pm$3.8) & ($\pm$1.4) & ($\pm$1.3)\\
    Supervised & 66.9 & 59.1 & 53.9 & 61.0  \\
    & ($\pm$2.2) & ($\pm$1.7) & ($\pm$1.5) & ($\pm$2.1) \\
    Human TA & 82.5 & -- & -- & -- \\
    \midrule
    &  \multicolumn{4}{c}{Held-out question} \\
    Model & AP & P@50 & P@75 & ROC-AUC \\
    \midrule
    ProtoTransformer & \textbf{88.1} & 89.8 & 82.1 & 89.0 \\
    & ($\pm$1.5) & ($\pm$2.1) & ($\pm$0.8) & ($\pm$0.9)\\
    Supervised & 64.5 & 58.3 & 53.6 & 59.5 \\
    & ($\pm$1.4) & ($\pm$1.6) & ($\pm$1.0) & ($\pm$1.6)\\
    Human TA & 82.5 & -- & -- & -- \\
    \midrule
    &  \multicolumn{4}{c}{Held-out exam} \\
    Model & AP & P@50 & P@75 & ROC-AUC \\
    \midrule
    ProtoTransformer & \textbf{74.2} & \textbf{77.3} & \textbf{67.3} & \textbf{77.0} \\
    & ($\pm$1.6) & ($\pm$2.7) & ($\pm$2.0) & ($\pm$1.4)  \\
    Supervised & 65.8 & 60.1  & 54.3 & 60.7  \\
    & ($\pm$ 2.1) & ($\pm$3.0) & ($\pm$1.8) & ($\pm$1.6) \\
    Human TA & 82.5 & -- & -- & -- \\
    \bottomrule
    \end{tabular}
    \caption{{\small Performance for feedback prediction in terms of precision and area under the ROC curve. P@X is the precision at a recall of X. Averaged over 3 runs.}}
    \vspace{-0.4cm}
    \label{table:main}
\end{wraptable}

\textbf{Computer Science Course Dataset}$\quad$
We use a corpus of student solutions to 4 final exams and 4 midterm exams from a  university
``introduction to computer science'' course collected over 3 years. Every student solution has feedback, in the form of a rubric, from at least one course teaching assistant. 10\% of the questions were annotated by more than one teaching assistant to measure grader consistency.
In total, the dataset includes 63 exam questions with 24.8k student solutions in Python. We remove rubric options with fewer than $K+Q$ positive examples and those where all students scored perfectly.
After processing, we have 259 tasks, of which 20 are held-out.
During meta-training, we randomly sample $K=10$ examples per class to form the support set $\mathcal{S}$ and $Q=10$ examples for the query set $\mathcal{Q}$.
The corpus also contains student solutions to 55 assignment questions.
However, assignment solutions are 8 times longer than exam solutions, as the two distributions are different.
On the next page, we investigate using assignments to construct additional tasks.


\textbf{Few-shot Feedback Tasks} Recall the feedback challenge from Section~\ref{sec:background}. To make this challenge amenable to few-shot learning, we need to build tasks from a rubric. The most obvious choice is to treat each rubric item as a task. However, this faces technical challenges.
First, options in a rubric item are very unbalanced e.g. some misunderstandings such as ``variable scoping issue'' are rare.
Second, rubric items have different numbers of options, requiring models to generalize over varying class counts.
We instead opted to \textit{treat each rubric option as a task}. For example, in Figure~\ref{fig:rubric}, the first rubric item on the left would comprise of four tasks. This implies that the same student programs may appear in multiple tasks, although the labels will be different. This also implies that all tasks are binary classification problems. Under this formulation, the number of tasks increases considerably, which we found to be crucial. In creating train test splits, we will always ensure rubric options from the same item are in the same split, preventing information leakage.

In addition, every programming question comes with a text description of the  prompt.
We propose to use both the question and the rubric item as side information to condition the embedding, as discussed in Section~\ref{sec:code}.
Concretely, let $z = (z_1, z_2, ..., z_T)$ be a sequence of tokens representing the prompt and $z' = (z'_1, z'_2, ..., z'_{T'})$ represent the rubric description.
Then we condition the embedding of any example $x_i$ in the support or query set on $g_\phi(z) + g_\phi(z')$, the sum of the prompt and rubric embeddings. In practice, we choose $g_\phi$ to be a pretrained SBERT.

\textbf{Evaluation Protocol}$\quad$ We evaluate the model on three splits: first, we reserve 10\% of rubric items, uniformly sampled, for meta-test tasks.
In this \textit{held-out rubric items} setup, the same student programs might appear in both splits, but under different rubric items. (We could imagine a setting where instructors partially grade every exam while the model is responsible for the rest.) Second, we hold-out a set of questions for meta-test split. This \textit{held-out question} setup ensures all rubric items for the same question are in the same split. (We could imagine instructors leaving a few questions to be auto-graded.)
Third, we hold-out a full exam. This \textit{held-out exam} setup poses a more difficult challenge as unseen tokens will appear at meta-test time. This more faithfully models a few-shot autonomous system.

\begin{wrapfigure}{r}{0.45\linewidth}
\centering
\includegraphics[width=\linewidth]{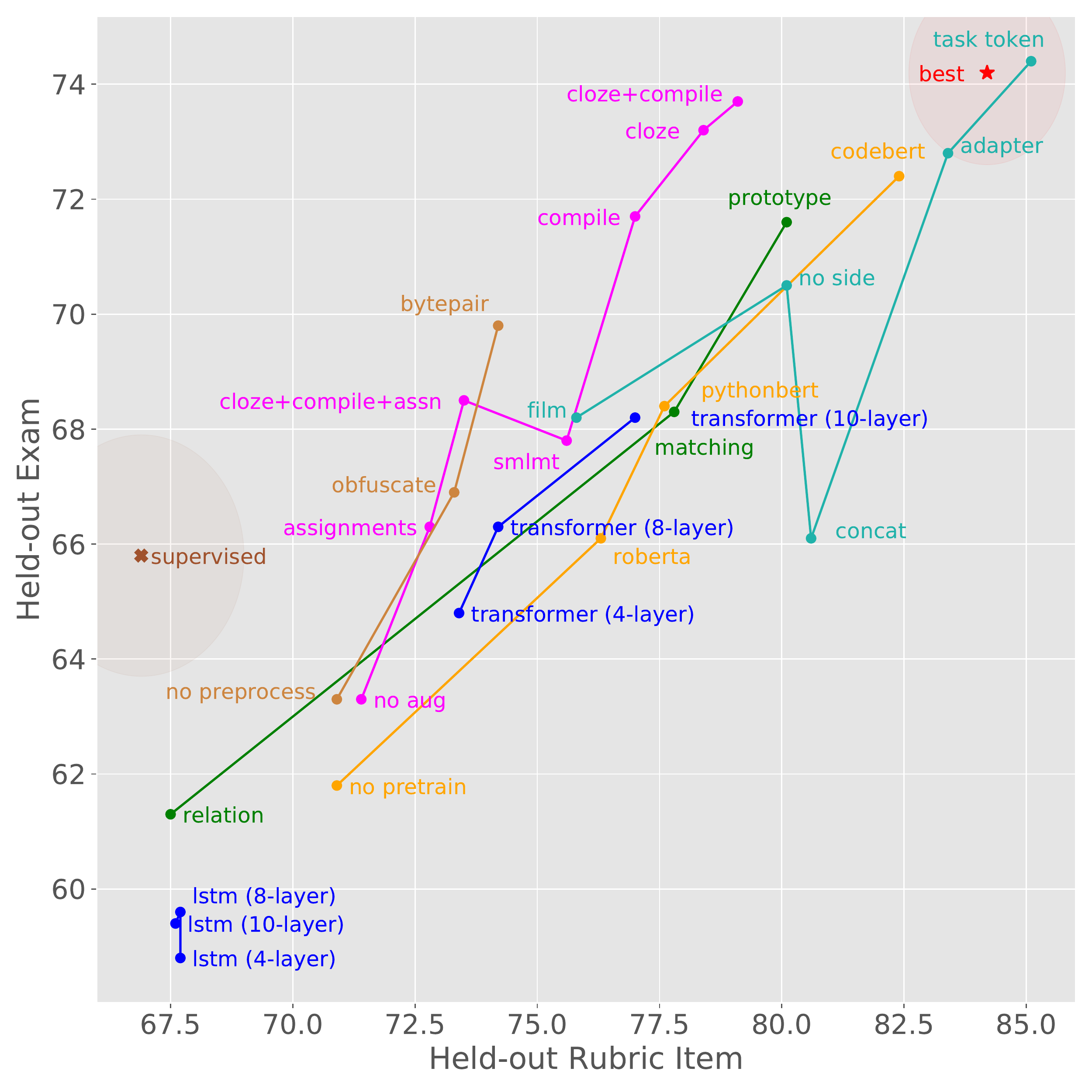}
\caption{{\small\textbf{Model Comparisons:} model variations compared to the best configuration (red). Shaded regions represent repeated runs. The x-axis plots AP on held-out rubric items whereas the y-axis plots held-out exams.}}
\label{fig:ablation}
\vspace{-2em}
\end{wrapfigure}

\textbf{Main Results}$\quad$
Table~\ref{table:main} reports the test performance of the ProtoTransformer on providing feedback to student solutions in both splits, averaged over three runs with different exams and rubrics being held-out.
We train the model for 300 epochs, preprocess code with BPE, and augment the meta-train set with 10\% cloze and 10\% compilation tasks.
See Figure~\ref{fig:model} for a model summary.
For metrics, we report AP, precision at recall 0.50 and 0.75, and area under the ROC curve (ROC-AUC).

We include a supervised baseline that fits a new model \textit{per task}. For every held-out task, we use $K$ examples to train a classifier from student code to a prediction for the rubric option associated with the task. We evaluate performance on $Q$ held-out examples.
To ensure the baseline is comparable, we use the same setup as the ProtoTransformer.
To train the supervised baseline, we found it crucial to use pretrained weights from CodeBERT, otherwise performance remained at chance.
Further, we estimate human performance (``human TA'' in Table~\ref{table:main}) by computing agreement in grading between multiple teaching assistants (TA) assigned to grade the same student solutions.
We found agreement to be relatively low at 82.5\%.
This variance can be attributed to multiple interpretations of student misconceptions from code as well as grader fatigue.
As the few-shot model is trained with labels from many TAs, in principle, a perfect model can outperform any individual grader.

Table~\ref{table:main} shows that the ProtoTransformer outperforms the supervised baseline by 17\% in held-out rubrics, 23\% in held-out questions, and 9\% in held-out exams.
Further, in held-out rubrics and questions, we find the ProtoTransformer to be highly performant, surpassing human precision by 1.7 points absolute and 5.6 points absolute, respectively.
The strong performance in the held-out questions setting might suggest a benefit of meta-training on all rubrics of a question at once.
As held-out exams pose a more difficult  challenge, we see a decrease of 10 points in AP and 5 points in ROC-AUC compared to  held-out rubrics.
However, the ProtoTransformer still outperforms the supervised baseline by a healthy margin, showing the utility of meta-learning.

\textbf{Model Comparisons}$\quad$
To pinpoint the effectiveness of individual techniques used in the ProtoTransformer, we conduct comparisons that change the architecture, auxiliary tasks, pretraining,  and more.\
To compare to prior methods, we vary the meta-learning algorithm.
Figure~\ref{fig:ablation} show the results.
In each comparison, all hyperparameters were kept constant aside from the one of interest.

Comparing the prototypical network to competing meta-learning algorithms, such as matching networks \citep{vinyals2016matching} and relation networks \citep{sung2018learning}, we find the prototypical network to outperform others by 3 and 12 points respectively (green curve).
We also find a significant benefit (of up to 10 points) in using transformer architectures opposed to recurrent architectures (blue curves). As the number of stacked transformer layers increased, the performance strengthened. For these depth experiments, models were trained from scratch.

We find that adding either the cloze or compile augmentation tasks introduced in Section~\ref{sec:methods} improves performance by 6-7 points (pink curve in Figure~\ref{fig:ablation}), with adding both further increasing performance by 1 point.
Notably, we find that adding cloze tasks improves performance by 3 points over adding SMLMT, despite the similarity between the two methods. Recall that SMLMT predominantly masks function and variable names, unlike cloze.
These results suggest that tasks predicting diverse naming tokens can negatively impact performance.
Finally, we investigate using student code from course assignments to augment the meta-training tasks.
Although these provide a 1 to 2 point benefit, it does not rival the performance of using cloze or compile tasks.
The much longer solutions from assignments are different in distribution, likely distracting the meta-learner.
We see evidence of this when using cloze, compile, and assignment tasks together: performance is 5 points worse than using cloze and compile tasks alone.

To measure the impact of unlabeled data, we compare several pretraining schemes. First, to confirm that unsupervised data has a positive impact, we compare performance using CodeBERT weights as initialization versus a random initialization (``no pretrain'' in the yellow curve).
We observe over 10 points of increase for both held-out rubrics and exams, suggesting pretraining to be crucial.
Moreover, to show that pretraining on unlabeled \textit{code} is important, we compare CodeBERT to the standard RoBERTa which is trained on natural language. Using RoBERTa weights shows roughly half of the improvement of CodeBERT over random initialization.
While CodeBERT is fit on a corpus of 6 programming languages (including Python), we finetuned a RoBERTa model only on the Python subset of CodeSearchNet \citep{husain2019codesearchnet}, 1.1M of the 6.5M total examples.
Using PythonBERT outperforms standard RoBERTa but does not match CodeBERT.

Incorporating side information proved useful (cyan curve).
We compared the method proposed in Section~\ref{sec:methods} (``Task Token'') to several alternatives.
FiLM \citep{perez2018film} uses side information to learn an additive and multiplicative shift applied in each transformer layer.
The ``adapter'' approach fits a bottleneck network to combine side information with attention outputs, inspired by \citep{houlsby2019parameter}.
Finally, the ``Concatenation'' approach merely joins side information with the final embedding post-transformer.
All in all, the ``Task Token'' approach is the simplest and most performant.

Lastly, to test preprocessing techniques, we compare the ProtoTransformer trained using obfuscated code, byte-pair encoded (BPE) code, and vanilla code (``no preprocess'').
While both obfuscation and BPE improve performance, the latter achieves 1 to 3 points of additional improvement (brown curve in Figure~\ref{fig:ablation}).
We hypothesis this to be because BPE preserves some semantic meaning in variable and function names (e.g. \texttt{sum = x + y}) whereas obfuscation does not.

%% file: sections/deployment.tex
\textbf{Real World Deployment}$\quad$
We deployed our approach to provide feedback to 16,000 student solutions on a diagnostic programming exam in a  course offered by a tier 1 university.
Normally, no feedback is given to students as it is not feasible for the teaching team to properly examine so many solutions. Rather, students are asked to self assess.
On an important ethical note, our model's feedback was used solely for enrichment, and had no impact on course grading or evaluation.
1,096 students responded to a survey after receiving 15,134 pieces of feedback. The students' reception to the feedback was overwhelmingly positive: Across the 15k pieces of feedback students agreed with AI suggestion 97.9\% $\pm$ 0.001 of the time. The 1,096 students were asked how useful they found the feedback on a 5 point scale. Their average rating was 4.6 $\pm$ 0.018 out of 5.
To the best of our knowledge this is the first successful deployment of AI feedback for open ended student code.
Analysis of the long term impact of such feedback, the reception of students to messaging concerning AI delivered feedback and the user interface for delivering such feedback is ongoing research.

%% file: sections/related.tex
\section{Related Work}

\paragraph{Meta-learning}
There is a rich collection of literature on meta-learning spanning multiple decades \citep{schmidhuber1999general,bengio1990learning,bengio1992optimization,younger2001meta,vanschoren2018meta,hospedales2020meta}. Our  model is based on the popular prototypical network \citep{snell2017prototypical}.
Traditionally, prototypical networks have focused on few-shot visual learning e.g. with Omniglot \citep{lake2011one} and miniImageNet \citep{vinyals2016matching}, although recent work has expanded to text classification \citep{sun2019hierarchical,gao2019hybrid,wu2019learning,geng2019induction,obamuyide2020model} and medical diagnosis \citep{prabhu2019few}.
In this vein, we study a new problem of few-shot code classification for education.
Recent research has expanded the prototypical network in several directions.
One such direction \citep{medina2020self,rajasegaran2020self} joins self-supervision together with few-shot learning.
Similar papers \citep{liu2019prototype,ren2018meta,bateni2020improving,liu2018learning} leverage unlabeled examples to modify prototype embeddings.
Our approach does not use unlabeled data at the few-shot learning stage, only for initialization.
Recent work also introduced task-specific parameters during meta-optimization \citep{oreshkin2018tadam,logeswaran2020few,denevi2020advantage}. Unlike this direction, our method has no trainable task parameters, only fixed side information.
On a third front, many papers have studied task curation:  AAL \citep{antoniou2019assume} randomly assigns labels; CACTUS \citep{hsu2018unsupervised} uses unsupervised clusters as labels; LASIUM \citep{khodadadeh2020unsupervised} fits a generative model to sample task examples.
In computer vision, \citep{liu2020task,khodadadeh2019unsupervised} create new tasks from image transformations.
In natural language, DReCa \citep{murtydreca} clusters BERT embeddings whereas SMLMT \citep{bansal2020self} predicts masked tokens for tasks.
Building on these, we curate tasks specific to code, which we find to work much better than tasks specifically for natural language.

\textbf{Machine learning for code } Applying neural networks to code (outside of education) has a substantial body of work \citep{allamanis2018survey}, studying type inference \citep{hellendoorn2018deep,pandi2020opttyper}, code summarization \citep{allamanis2016convolutional,iyer2016summarizing}, or program induction \citep{kant2018recent,devlin2017neural,huang2018natural}. Several works investigate learning code representations \citep{alon2018code2seq,feng2020codebert,kanade2019pre,jain2020contrastive}, which we leverage to initialize our few-shot model. However, on their own, we find that existing representations out-of-the-box are not sufficient for feedback.

\textbf{Machine learning for education } In knowledge tracing, a machine models student knowledge as they interact with coursework \citep{corbett1994knowledge,piech2015deep,shin2020saint+}. However, these models are not able to provide feedback.
Recently, \citep{wu2019zero,malik2019generative} propose to provide feedback using expert-built grammars. However, developing faithful grammars becomes intractable for complex programs, such as university level exams.
Our approach relies only on standard annotation for a small set of examples: a cheap cost for any expert. We find good performance on university level student code.


%% file: sections/conclusion.tex
\section{Limitations and  Conclusion}
\label{sec:limit}
While promising, there are important limitations to consider. The few-shot learning setup implicitly assumes access to $K$ labeled examples per class. But in practice, we may have to label many more as some classes are rare. Table~\ref{table:degradation} shows how using less than $K$ annotations impacts performance. Second, it is not clear how far the approach can generalize e.g. a new exam or a new course.
On a broader scale, we are conscious of the responsibility of AI in education. While an automated system can be powerful, if it is not interpretable and equitable in its predictions, there is a potential negative impact to students' education. There could be bias especially if the model performs differently for lower performing students who have more complex solutions, raising ethical considerations. We believe it is crucial that our system works together with a teaching staff. In deployment, we took extreme care in quality analysis and crafting careful language around presenting feedback.

In summary, we introduced a few-shot approach to predict  feedback on student programming code, matching human experts in accuracy.
We observe the combination of transformers, side information, and augmented tasks made impactful improvements in education and NLP. As priori methods struggled with university assignments, we believe our work takes an important step forward.

%% file: sections/appendix.tex
\newpage
\appendix

\section{Model Details}

\subsection{Details: Unsupervised Pretraining}

Pretrained RoBERTa weights are taken from the HuggingFace Transformers repository \citep{wolf2019huggingface} under the key \texttt{roberta-base}. Pretrained CodeBERT \citep{feng2020codebert} weights are publically available at \url{https://github.com/microsoft/CodeBERT} and can be accessed with the HuggingFace Transformers library under the key \texttt{microsoft/codebert-base}. Pretrained PythonBERT weights are obtained by finetuning the top 6 layers of a pretrained RoBERTa model (\texttt{roberta-base}) on the CodeSearchNet repository \citep{husain2019codesearchnet} limited to Python code, unlike CodeBERT which is trained on multiple languages. We optimize with Adam with weight decay of 0.01, $\beta_1$=0.9, $\beta_2$=0.999 for 10 epochs, a batch size of 8, and the learning schedule from \citep{liu2019roberta} with a peak learning rate of 1e-4 and 10k warmup steps.

\subsection{Details: Task Augmentation}
All programs in the dataset are in python. For the ``compile'' task, we use the built-in python function \texttt{compile}. We use a \texttt{try-catch} statement on failure with the exception string denoting a class label. A successful compilation indices a ``success'' class label. We group compilation errors that contain the text ``Did you mean to print ...'' into one class, and similarly for the text ``invalid character in identifier ...''. In total, we have 26 possible compilation class labels, of which two are randomly chosen at a time to construct a task.

The SMLMT \citep{bansal2020self} and our ``cloze'' task are very similar as both create classification tasks that involve predicting the identity of masked tokens in a discrete sequence.
The primary difference for the cloze task is (1) it is always a binary classification problem, and (2) we ignore all special tokens (e.g. start of sentence, etc) and all function and variable names. As such, the masked tokens largely represent built-in python functions (e.g. \texttt{def} or \texttt{in}).

\subsection{Details: Side Information Architectures}
\label{sec:app:side}

\begin{figure}[h!]
\centering
\begin{subfigure}[b]{0.24\linewidth}
    \centering
    \includegraphics[width=0.65\textwidth]{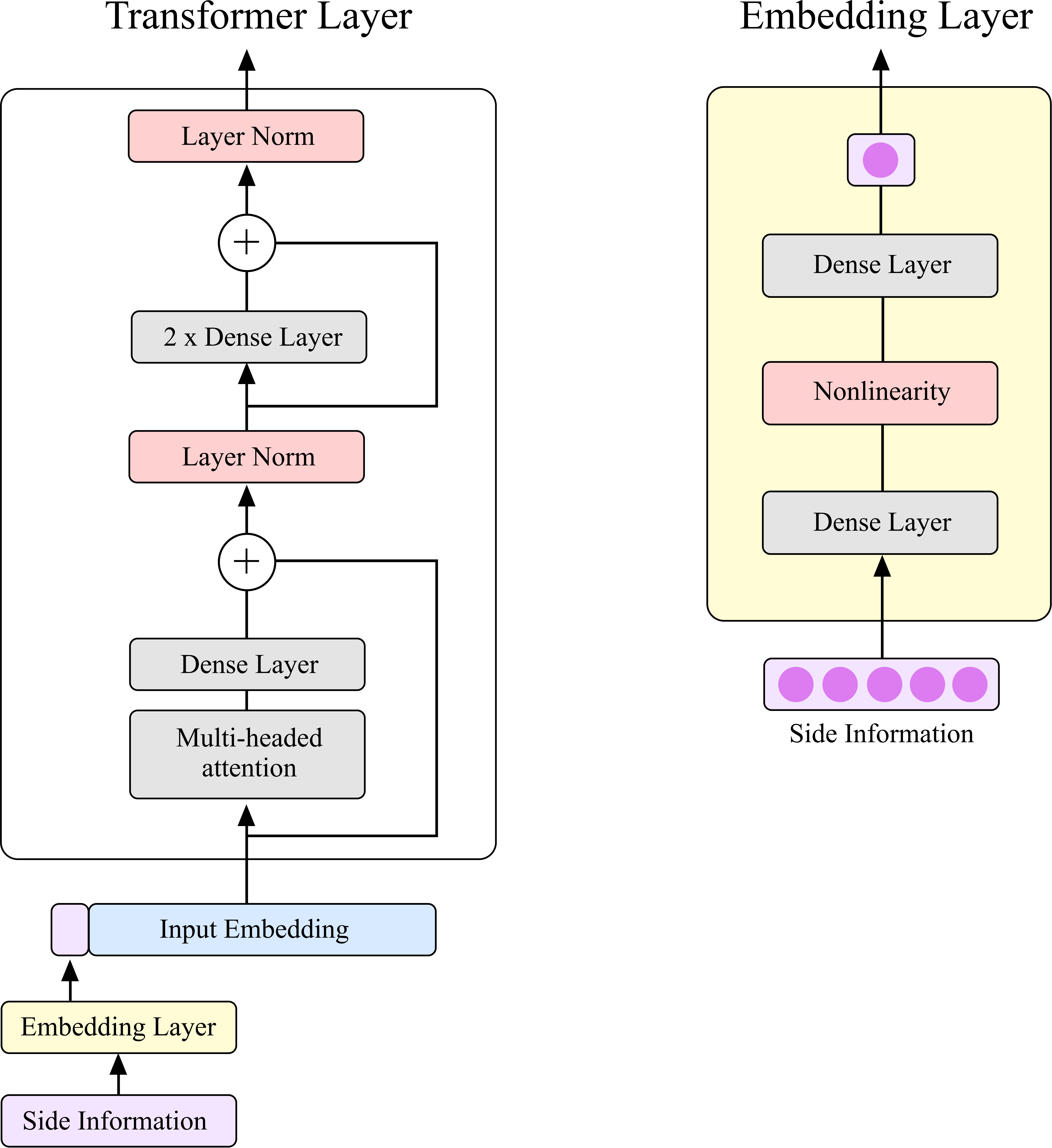}
    \caption{Task Token}
\end{subfigure}
\begin{subfigure}[b]{0.24\linewidth}
    \centering
    \includegraphics[width=0.8\textwidth]{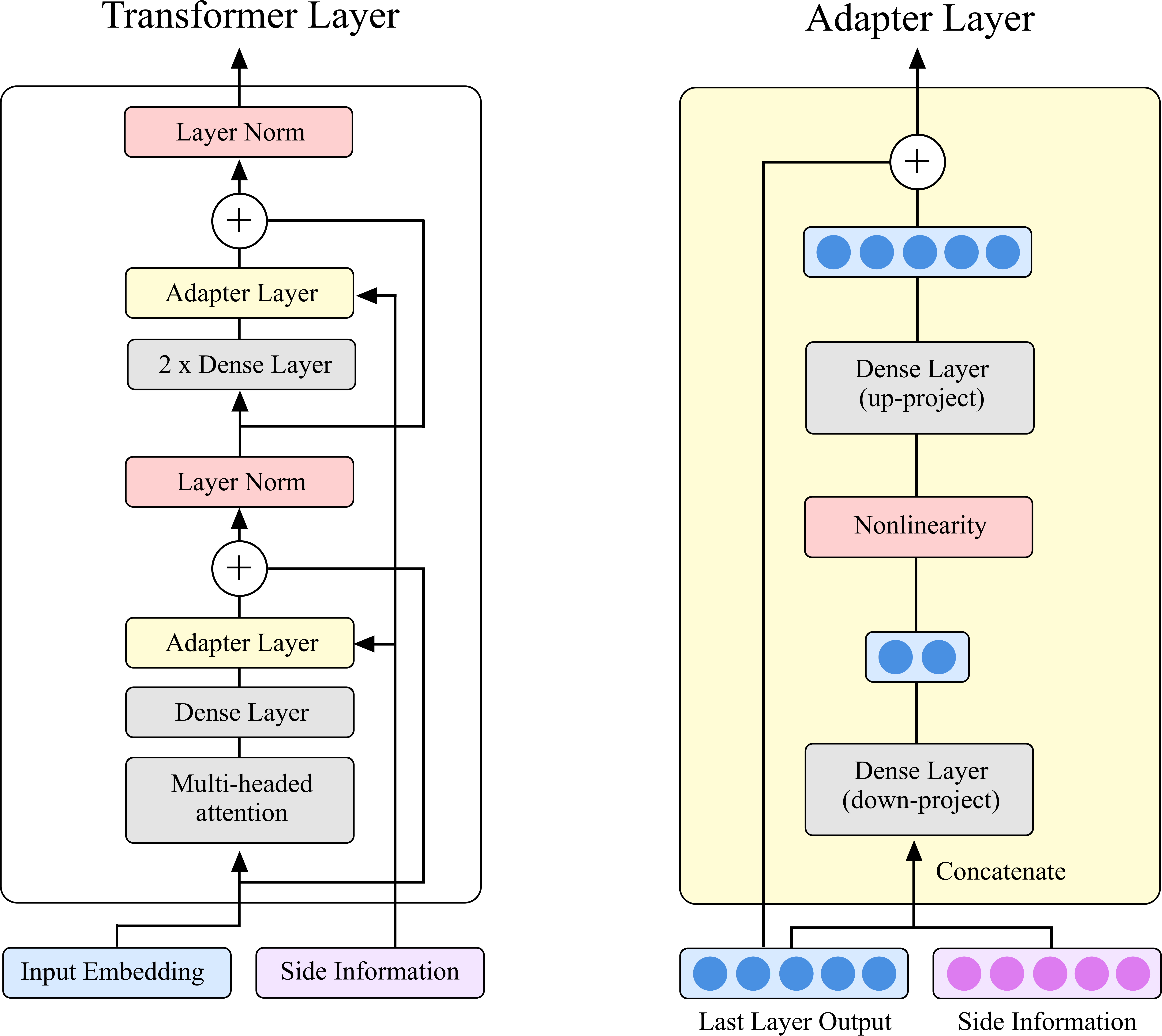}
    \caption{Adapter Network}
\end{subfigure}
\begin{subfigure}[b]{0.24\linewidth}
    \centering
    \includegraphics[width=0.8\textwidth]{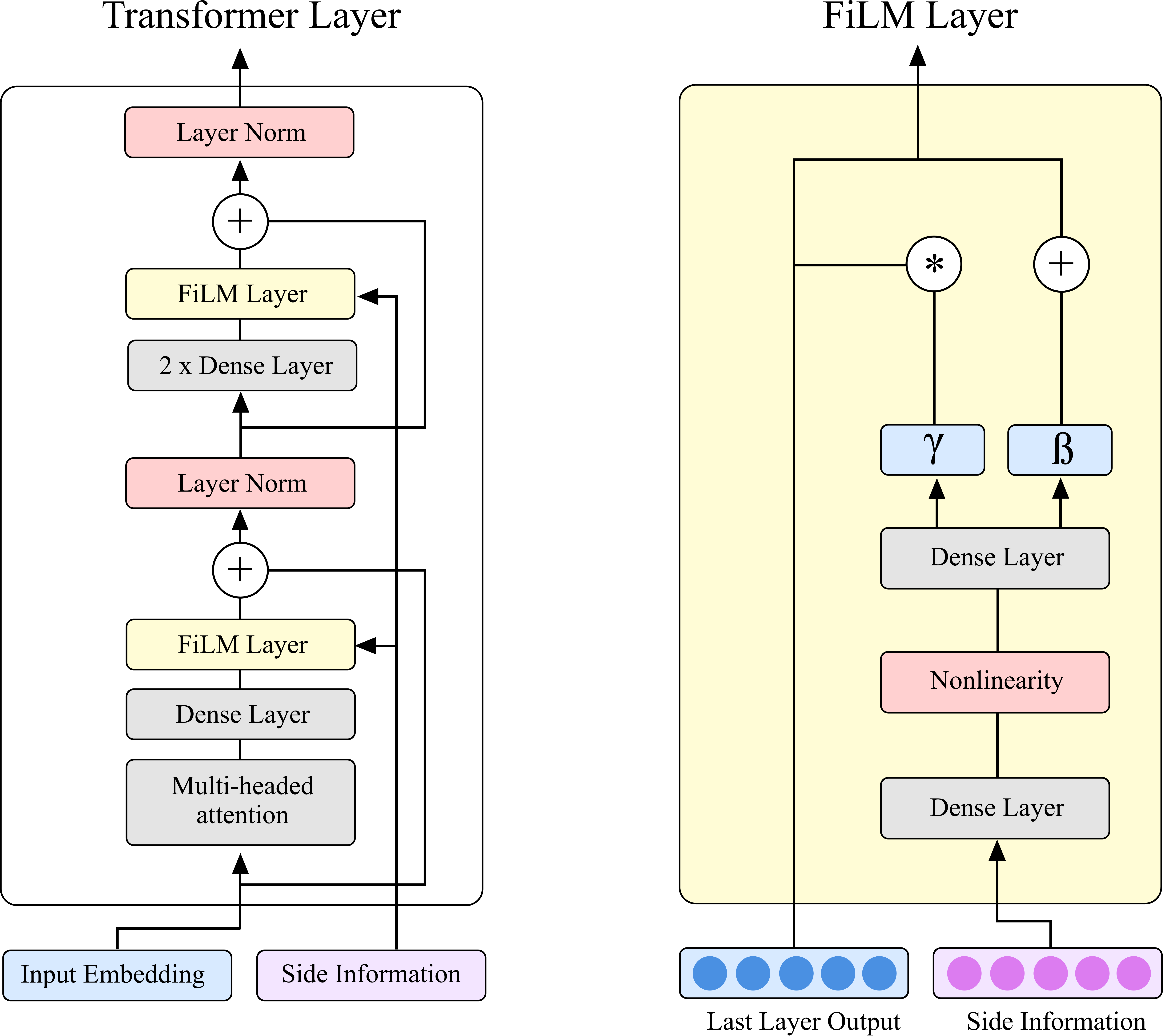}
    \caption{FiLM Network}
\end{subfigure}
\begin{subfigure}[b]{0.24\linewidth}
    \centering
    \includegraphics[width=0.45\textwidth]{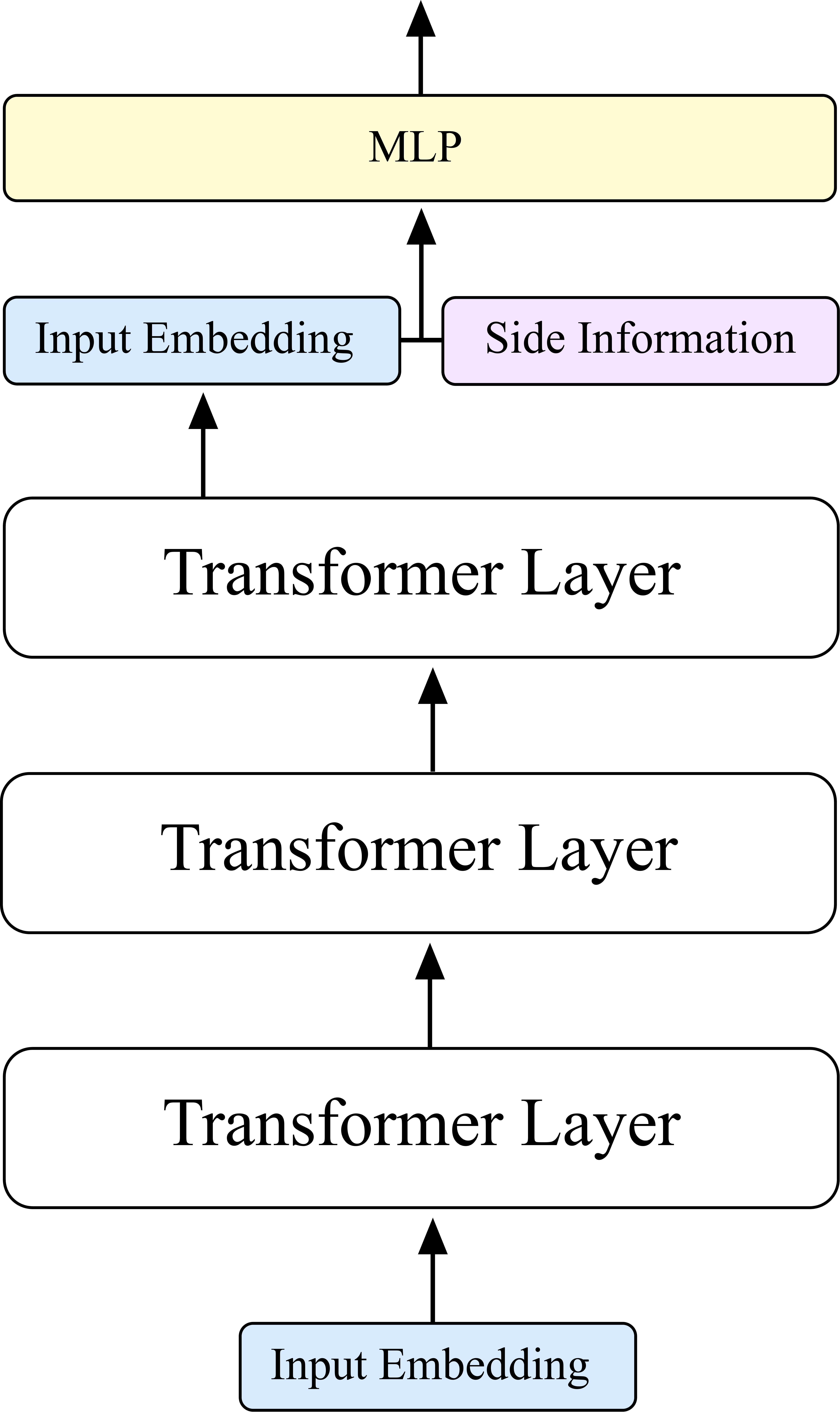}
    \caption{Concatenation}
\end{subfigure}
\caption{Four options for integrating side information into a stacked transformer architecture. Some approaches modify the input to the ProtoTransfer (a) while others modify each transformer layer (b,c), and others still modify the final transformer representation.}
\label{fig:app:arch}
\end{figure}

To incorporate side information in to the ProtoTransformer architecture, we considered four different options:

The simplest method, denoted ``concat'', first separately feeds the input sequence through the stacked transformer model, and computes an average embedding over non-padding sequence tokens of shape \texttt{batch\_size x dim}. This embedding is then concatenated with the side information, which we assume to be a vector, or equivalently, pre-embedded into a vector. The concatenated vector is then fed into an MLP to collapse the embedding back to  shape \texttt{batch\_size x dim}.
While simple, one disadvantage of this approach is that the side information does not directly influence how the input sequence is embedded in the transformer layers.

The second method, denoted ``FiLM'', attempts to modify the internal architecture of each transformer layer.
As shown in Fig.~\ref{fig:app:arch}c, two FiLM layers are added after dense layers in the transformer layer architecture.
The FiLM layer \citep{perez2018film} was proposed as a general method to condition convolutional neural networks on known information by learning and additive and multiplicative shift applied to the output of a layer in the neural network.
The FiLM idea was extended to transformers in TADAM \citep{oreshkin2018tadam}.
In this work, we propose to use FiLM to condition the embedding function on side information.
Side information is put through an MLP to produce two terms, $\gamma$ and $\beta$ which are then respectively multiplied and added (elementwise) to the output of the a dense layer in the transformer.

The third method, denoted ``adapter'', similarly edits the transformer layer architecture.
Proposed in \citep{houlsby2019parameter}, it was introduced as a parameter-efficient finetuning procedure.
Two ``adapter networks'' are added after dense layers in the transformer (see Fig.~\ref{fig:app:arch}b).
Each adapter is a bottleneck network that down-projects the input to a small hidden dimension before up-projecting it back to the original dimensionality, with an nonlinearity in between.
There is a residual connection between the original input and the reconstructed input of the adapter network.
The original architecture from \citep{houlsby2019parameter} did not include side information.
However, we propose to concatenate side information with the input (call this $X$) to the adapter network.
In this new design, the up-projection layer produces a vector of the same dimensionality as $X$ (not including added dimensions by concatenating side information).
The output of this adapter network is again the original $X$ with the reconstructed residual added. Intuitively, the residual captures side information.

The fourth and final approach, is based on TAM \citep{logeswaran2020few}. We use an MLP to project the side information to the side dimensionality as the token embedding in the transformer (e.g. 768 for RoBERTa). We then treat this vector as a ``special'' token, prepending it to the sequence of embedded tokens. See Fig.~\ref{fig:app:arch}a for an illustration.

\subsection{Details: Code Preprocessing}

For byte-pair encoding, we use the \texttt{RobertaTokenizer} from HuggingFace's Transformers repository.
To do obfuscation, we first must identify variable and function names.
To accomplish this, we use the \texttt{pythonlang.tokenize} function, which is a lexical scanner for python source code and semantically tags tokens with types (e.g. variable, function, etc.). This lexical scanner does not require the program to compile.
Then we denote a finite set of reserved tokens for representation token and variable names e.g. \texttt{<VAR:0>}, ..., \texttt{<VAR:100>}, or \texttt{<FUNC:0>}, ..., \texttt{<FUNC:10>}. When obfuscating program, every time we see a variable or function token, we replace all instances of it with the next available reserved token.
During training, we randomly shuffle variable and function tokens, e.g. all instances of \texttt{<VAR:0>} to \texttt{<VAR:5>}, in order to prevent memorizing.
The lexical scanner is also used for the ``cloze'' task to decide which tokens to mask.

\subsection{Details: Supervised Baseline}
The supervised baseline is trained for 25 epochs to avoid overfitting.
We first tried to train for 300 epochs to be a faithful comparison to the ProtoTransformer but found no improvements after 25 epochs.
Initializing weights from CodeBERT slowed overfitting to some extent. Without it, performance for the supervised baseline plateaued at chance (AP near 0.5).

\subsection{Details: Additional Examples}

\begin{figure}[h!]
\centering
\begin{subfigure}[b]{\textwidth}
    \centering
    \includegraphics[width=\linewidth]{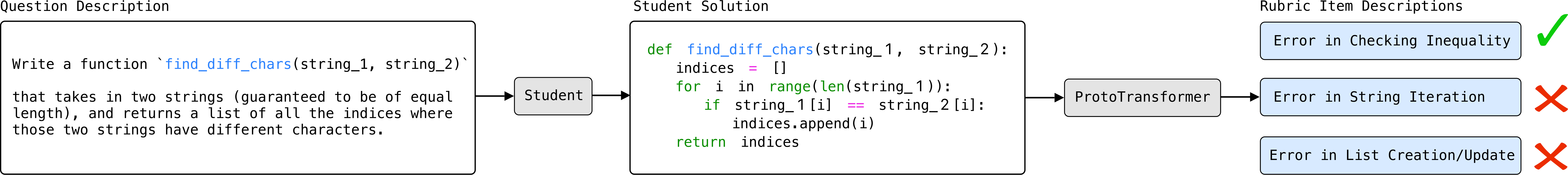}
    \caption{Example: string comparisons}
\end{subfigure}
\begin{subfigure}[b]{\textwidth}
    \centering
    \includegraphics[width=\linewidth]{./images/examples2.pdf}
    \caption{Example: file reading and processing}
\end{subfigure}
\begin{subfigure}[b]{\textwidth}
    \centering
    \includegraphics[width=\linewidth]{./images/examples3.pdf}
    \caption{Example: dictionary comparisons}
\end{subfigure}
\caption{\textbf{More Examples} of rubric items from a trained ProtoTransformer. The question description and the true misconception are shown in the left two boxes whereas the predicted rubric item is shown with a checkmark out of three possible items.}
\label{fig:examples:more}
\end{figure}

We provide more examples of ProtoTransformer predictions for feedback on exam questions. Fig.~\ref{fig:examples:more} shows two more examples in addition to the one in the main text.
Since we are not able to share student work, the examples are generated by an author who purposefully introduces a misconception in their solution.
In particular, in Fig.~\ref{fig:examples:more}a, the comparision betweens \texttt{string\_1} and \texttt{string\_2} should be an inequality. In Fig.~\ref{fig:examples:more}b, the line \texttt{if content[i][j] // 2 == 0:} should instead be \texttt{if content[i][j] \% 2 == 0:} e.g. replace integer division with a mod.
In Fig.~\ref{fig:examples:more}c, we did not check to see if the key \texttt{k} exists in the object \texttt{dict2} before accessing it.
In each of the cases, the ProtoTransformer predicts feedback labels that reflect the original misconception.

\section{Dataset Details}
\label{sec:supp:data}

\begin{figure}[h!]
\begin{subfigure}[b]{0.24\linewidth}
    \includegraphics[width=\textwidth]{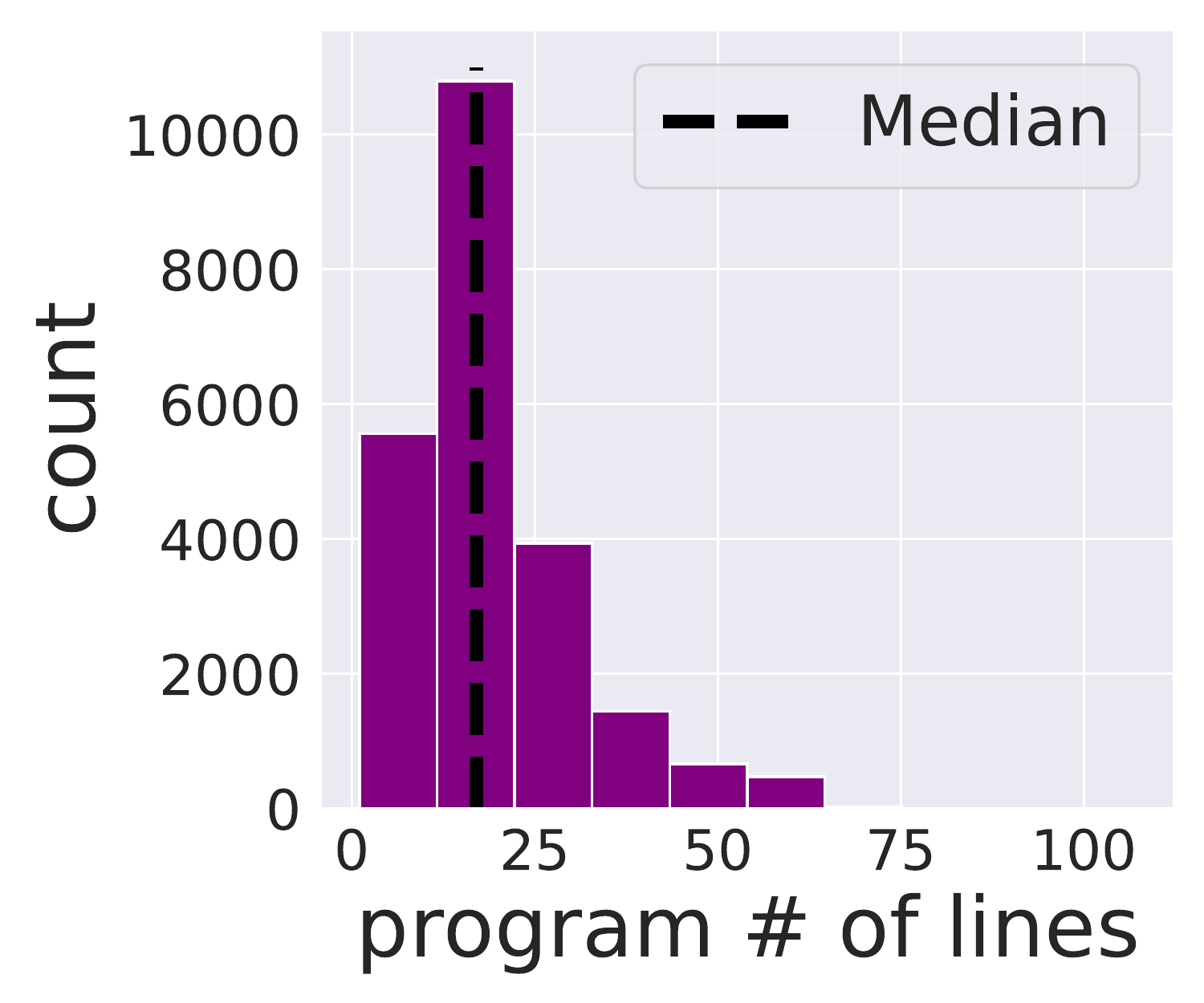}
    \caption{Exams}
\end{subfigure}
\begin{subfigure}[b]{0.24\linewidth}
    \includegraphics[width=\textwidth]{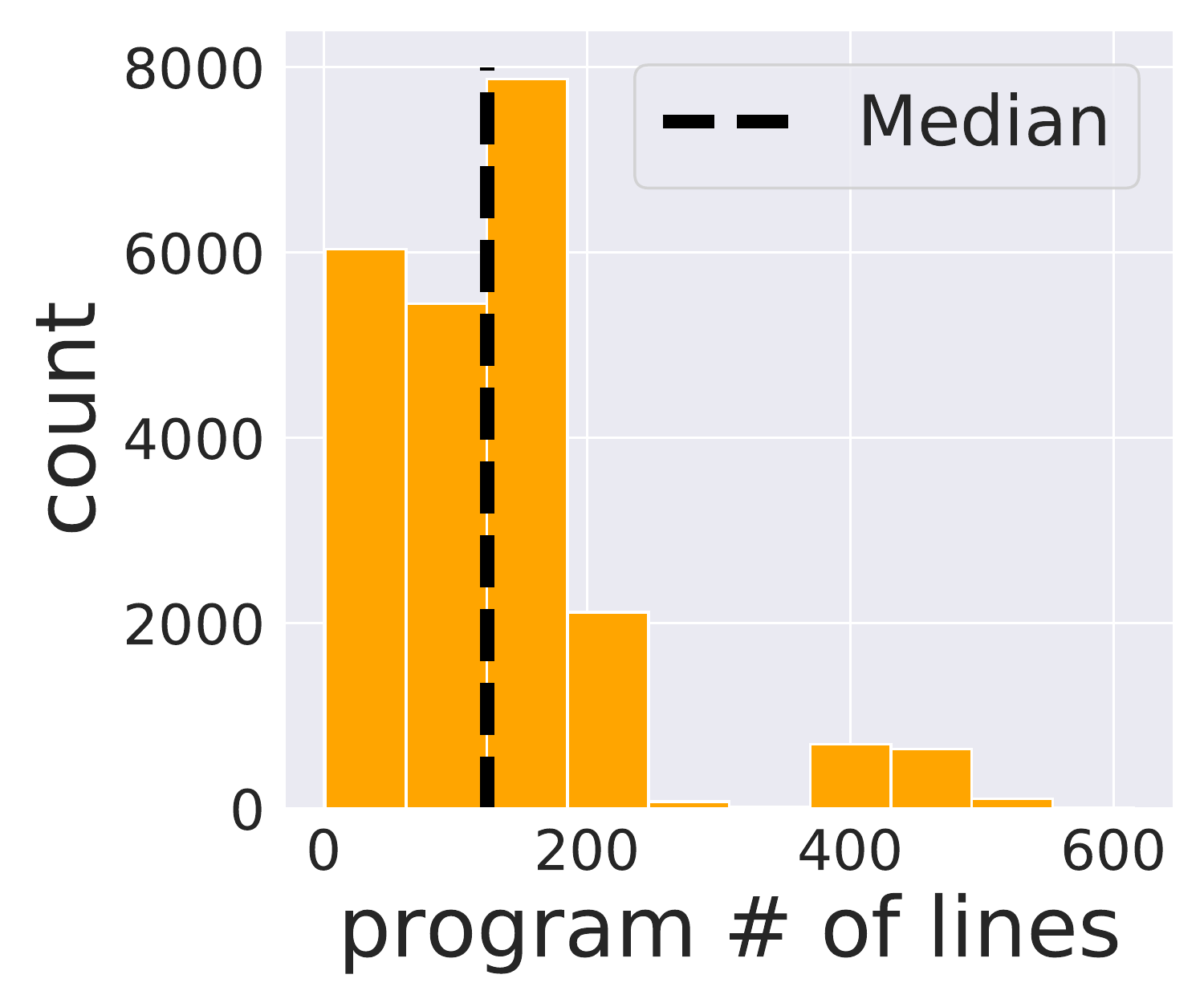}
    \caption{Assignments}
\end{subfigure}
\begin{subfigure}[b]{0.24\linewidth}
    \includegraphics[width=\textwidth]{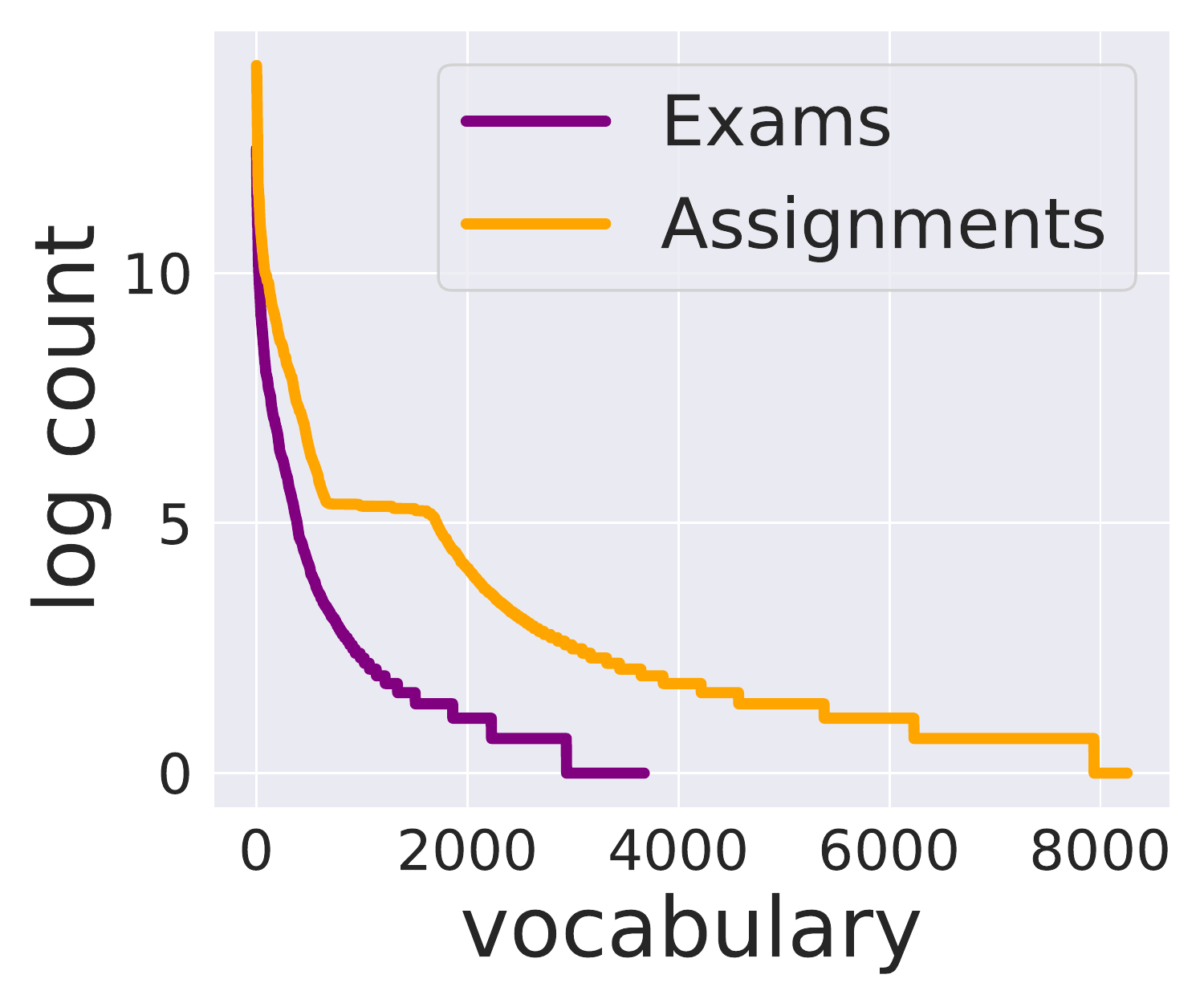}
    \caption{Token Uniqueness}
\end{subfigure}
\begin{subfigure}[b]{0.24\linewidth}
    \includegraphics[width=\textwidth]{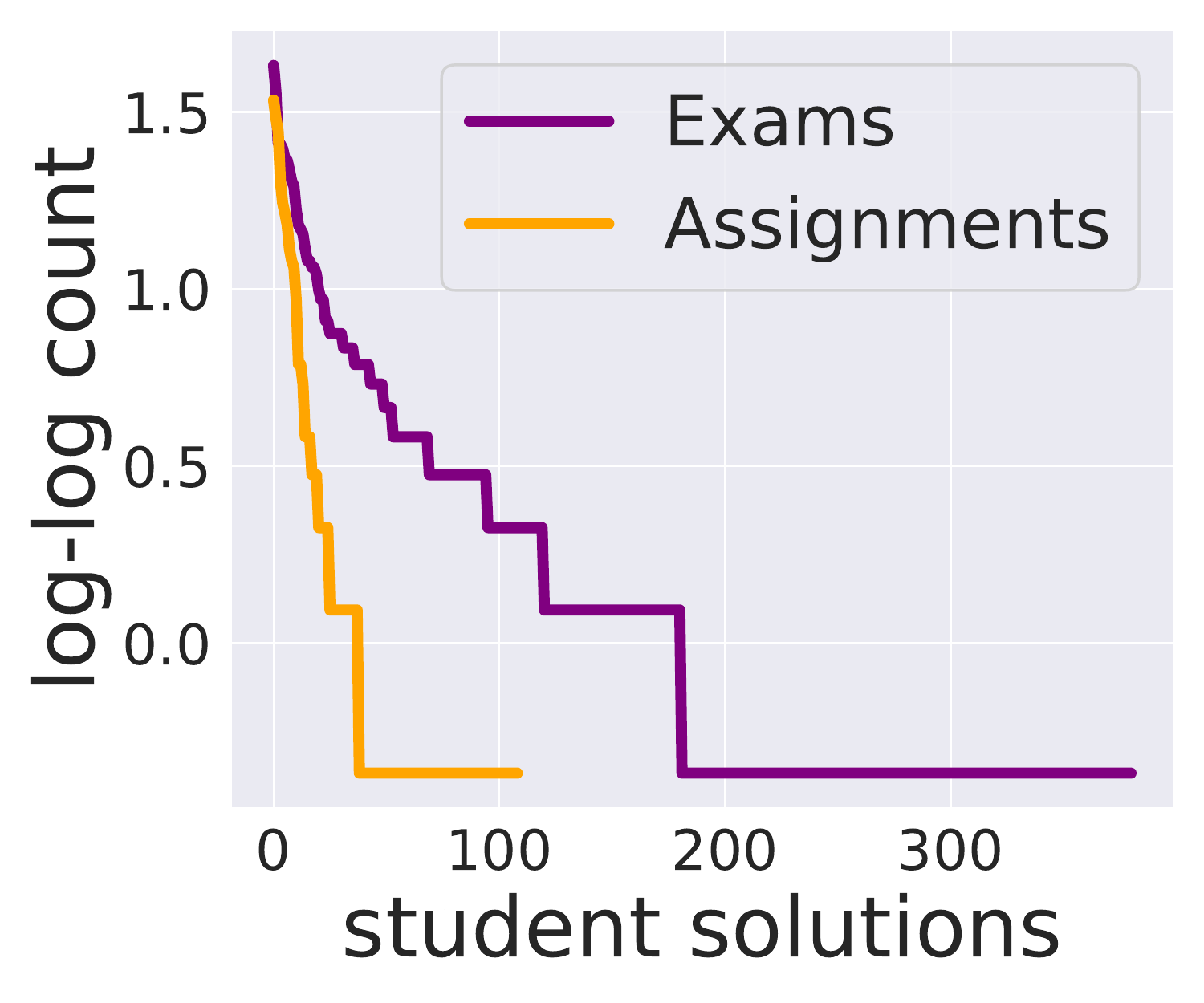}
    \caption{Program Uniqueness}
\end{subfigure}
\caption{Sub-figures (a) and (b) show number of lines in the exams and assignments (black dashed line shows median). Sub-figure (c) shows the log counts of unique tokens in exams and assignments.  Sub-figure (d) shows the log-log counts of unique programs.}
\label{fig:data_analysis}
\end{figure}

Fig.~\ref{fig:data_analysis} compares student solutions to course exams to course assignments in terms of code length and uniqueness.
Fig.~\ref{fig:data_analysis}a and b show that assignments are nearly 10 times longer than exams, as assignments are often more difficult than exam problems.
As a consequence, student solutions to assignment problems are much more likely to be unique as well (see Fig.~\ref{fig:data_analysis}d). The diversity of student solutions also results in a larger vocabulary size (see Fig.~\ref{fig:data_analysis}c). For purposes of learning algorithms, the distribution of student exam solutions and assignment solutions are significantly different.
It is likely much more difficult to model  assignment solutions.
A final note: our data analysis indicates that students programming code has a very long tail with a Zipf-like nature.
Fig.~\ref{fig:data_analysis} plots unique student solutions (x-axis) against the log-log count of number of appearances in the dataset (we log twice in order to visualize the fall-off).
It is clear that a handful of programs appear often whereas the majority of student solutions have not been seen before.
This property of student data makes traditional supervised learning approaches difficult as one has to annotate the tail to generalize.

\subsection{Dataset Specifics}
\label{sec:app:data:spec}

Student submissions were collected from 3 years of an introductory programming course at a university (information redacted for anonymity). Each course contains 500 to 1000 students. For every course, students are given programming assignments and two exams: one midterm and one final. We do not record student attempts, only the code of their final submission. A team of teaching assistants was assigned to grade every student submission according to a rubric, which we use as side information in the model. Because grading styles are not normalized, a set of randomly chosen problems are graded by multiple teaching assistants, which we use to compute precision.

Personal information about students was scrubbed from the dataset: we do not utilize grader or student information in the model, only the question text, rubric text, and solution. In the solution, all comments were removed.

\subsection{Dataset Release}
We are cognizant that the ``Introduction to computer science'' dataset of student solutions from Section~\ref{sec:dataset} is not public. We have attempted to open-source this dataset many times but found it extremely difficult to do so given important privacy concerns over student data. We are working towards open-sourcing a version of a dataset of student code for feedback prediction so that others can improve upon our approach. Our motivation on exploring applications of the ProtoTransformer to natural language was to construct a proxy task for others to compare against our method while we work towards a public dataset of student code.

\section{Additional Results}

\begin{figure}[h!]
\centering
\begin{subfigure}[b]{0.45\linewidth}
    \centering
    \includegraphics[width=0.75\linewidth]{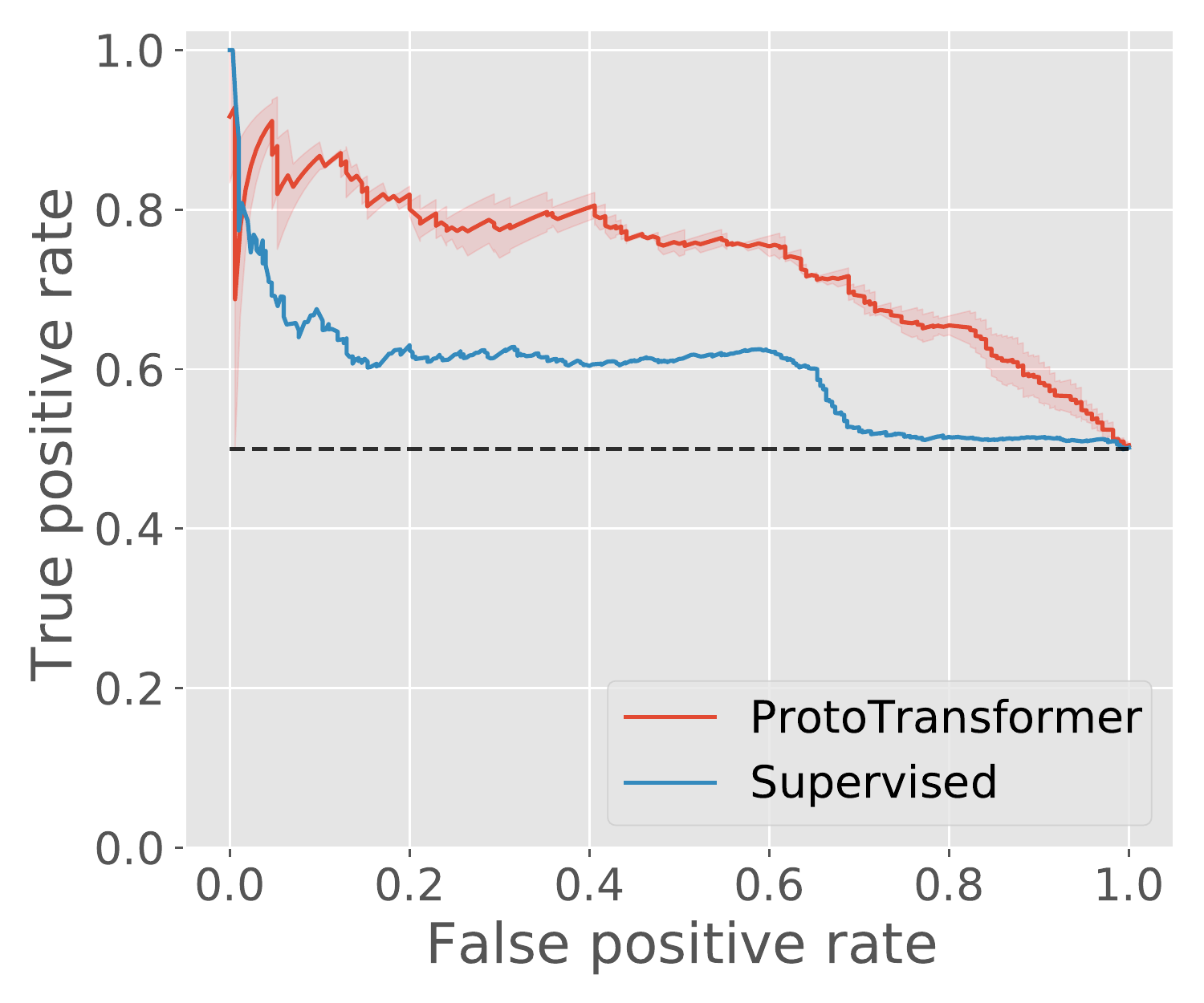}
    \caption{Precision-Recall}
\end{subfigure}
\begin{subfigure}[b]{0.45\linewidth}
    \centering
    \includegraphics[width=0.75\linewidth]{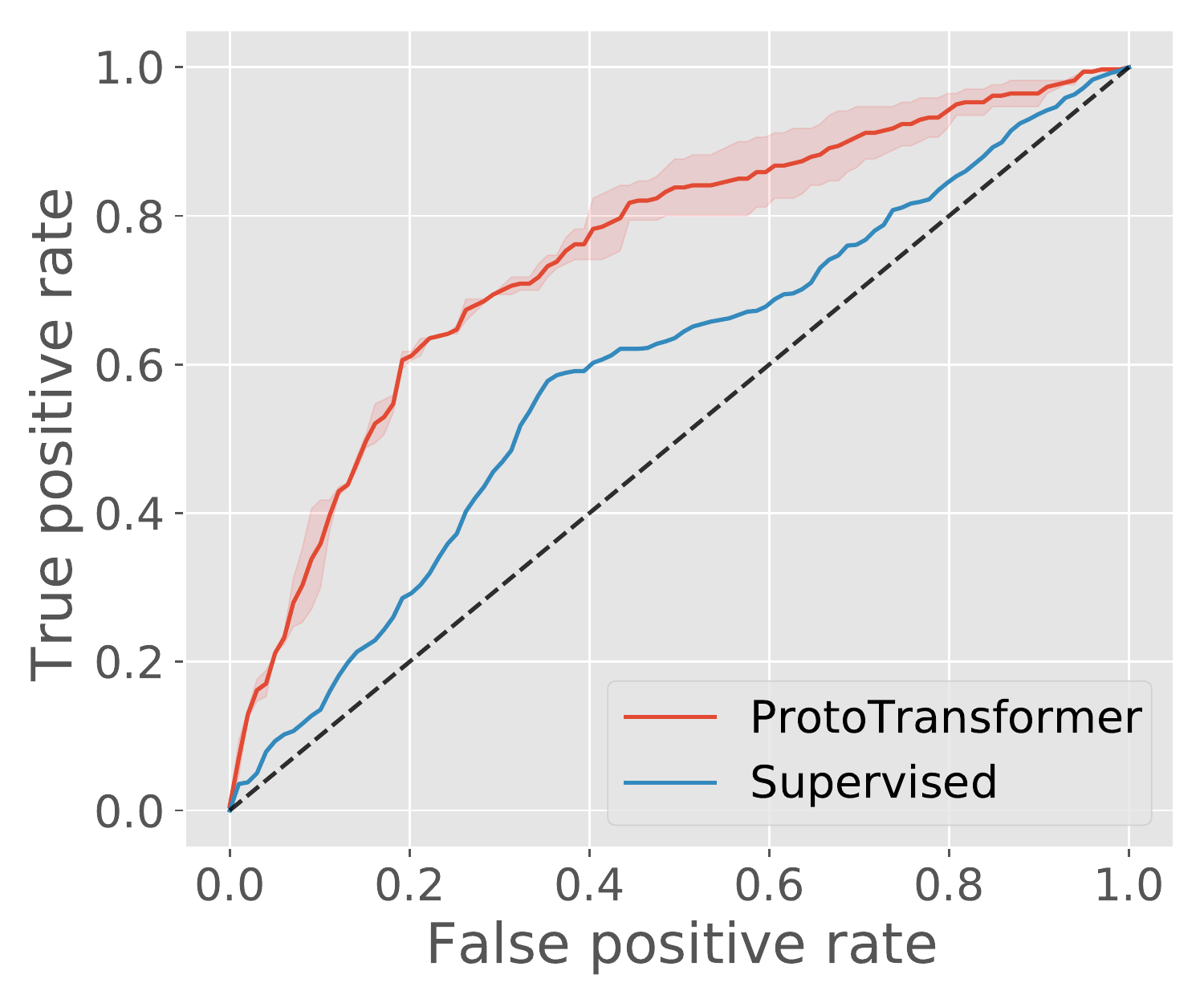}
    \caption{ROC curve}
\end{subfigure}
\caption{Comparision of the ProtoTransformer (red) to a supervised baseline (blue). The dotted black line represents chance.}
\label{fig:results}
\end{figure}

We include plots of the precision recall curve and ROC curve of the ProtoTransformer and supervised baseline. See Fig.~\ref{fig:results}.
The black dotted line represents performance of random chance. Thus, in both subfigures, a curve closer to the black dotted line is a worse performing model.

Next, we explore how meta-test performance degrades as we build prototypes in test time with fewer than $K$ examples (Note we will assume access to $K$ examples in meta-train tasks). Table~\ref{table:degradation} shows a gradual loss in performance as fewer shots are provided during meta-test. However, the performance drop is relatively small when using 5 shots (half the number).

\begin{table}[h!]
    \small
    \centering
    \begin{tabular}{lcccc}
        \toprule
        & \multicolumn{2}{c}{Held-out rubrics} & \multicolumn{2}{c}{Held-out exams}  \\
        $K$ shots & AP & ROC-AUC & AP & ROC-AUC \\
        \midrule
        10 & $84.2$ & $82.9$ & $74.4$ & $77.1$ \\
        5 & $82.1$ & $80.8$ & $72.9$ & $73.6$ \\
        2 & $77.2$ & $76.6$ & $69.7$ & $70.2$ \\
        1 & $66.1$ & $69.6$ & $58.1$ & $62.0$ \\
        \bottomrule
        \end{tabular}
    \caption{\textbf{Effect of support set size on meta-test performance}: Using a ProtoTransformer Network trained on meta-training tasks with $10$ shots, we find gradual loss in performance as fewer shots are provided during meta-test. }
    \label{table:degradation}
\end{table}

\section{Additional Ablations}
\label{sec:app:ablation}
We include two ablation experiments not shown in the main paper.
First, we vary the amount of task augmentation to answer the question: \textit{how many synthetic tasks is too many?}
From Table~\ref{table:app:ablation1}, adding 10\% of augmented tasks results in the best performing model. As the percentage of synthetic tasks grows below or above 10\%, performs continues to decrease.
Second, we want to know \textit{how does the number of shots affect performance?} In our main experiments, we set $K=10$.
As Table~\ref{table:app:ablation2} shows, the bigger $K$ is, the higher the performance, though there may be a marginally decreasing effect.
In practice, we cannot choose $K$ to be too big. For the education application, it is not feasible to annotate too many examples per feedback class. By choosing $K=10$, we believe it to be a compromise between annotation effort and performance.
In Table~\ref{table:app:ablation1}, we also include numbers for the ablations in the main text.

\begin{table}[h!]
    \tiny
    \centering
    \begin{subtable}{.45\linewidth}
        \centering
        \begin{tabular}{lcc}
        \toprule
        & \multicolumn{2}{c}{AP}  \\
        Meta Algorithm & Held-out rubric & Held-out exam \\
        \midrule
        Prototypical Network & \textbf{80.1} & \textbf{71.6} \\
        Matching Network & 77.8 & 68.3 \\
        Relation Network & 67.5 & 61.3 \\
        \bottomrule
        \end{tabular}
        \caption{\textbf{Ablation:} meta-learning algorithm.}
    \end{subtable}
    \begin{subtable}{.45\linewidth}
        \centering
        \begin{tabular}{lcc}
        \toprule
        Architecture & Held-out rubric & Held-out exam \\
        \midrule
        LSTM (4x) & 67.7 & 58.8 \\
        LSTM (8x) & 67.7 & 59.6 \\
        LSTM (10x) & 67.6 & 59.4\\
        Transformer (4x) & 73.4 & 64.8 \\
        Transformer (8x) & 74.2 & 66.3 \\
        Transformer (10x) & \textbf{77.0} & \textbf{68.2} \\
        \bottomrule
        \end{tabular}
        \caption{\textbf{Ablation:} embedding network architecture.}
    \end{subtable}
    \begin{subtable}{.45\linewidth}
        \centering
        \begin{tabular}{lcc}
        \toprule
        Augmented Task & Held-out rubric & Held-out exam \\
        \midrule
        None & 71.4 & 63.3 \\
        SMLMT & 75.6 & 67.8 \\
        Assignments (Assn.) & 72.8 & 66.3 \\
        Compile & 77.0 & 71.7 \\
        Cloze & 78.4 & 73.2 \\
        Cloze, Compile & \textbf{79.1} & \textbf{73.7} \\
        Cloze, Compile, Assn. & 73.5 & 68.5 \\
        \bottomrule
        \end{tabular}
        \caption{\textbf{Ablation:} task augmentation.}
    \end{subtable}
    \begin{subtable}{.45\linewidth}
        \centering
        \begin{tabular}{lcc}
        \toprule
        Pretraining & Held-out rubric & Held-out exam \\
        \midrule
        None & 70.9 & 61.8 \\
        RoBERTa & 76.3 & 66.1 \\
        CodeBERT & \textbf{82.4} & \textbf{72.4} \\
        PythonBERT & 77.6 & 68.4 \\
        \bottomrule
        \end{tabular}
        \caption{\textbf{Ablation:} pretraining scheme}
    \end{subtable}
    \begin{subtable}{.45\linewidth}
        \centering
        \begin{tabular}{lcc}
        \toprule
        Side Info. Arch. & Held-out rubric & Held-out exam \\
        \midrule
        None & 80.1 & 70.5 \\
        Concatenation & 80.6 & 66.1 \\
        Adapter & 83.4 & 72.8 \\
        Task Token & \textbf{85.1} & \textbf{74.4} \\
        FiLM & 75.8 & 68.2 \\
        \bottomrule
        \end{tabular}
        \caption{\textbf{Ablation:} side information architecture.}
    \end{subtable}
    \begin{subtable}{.45\linewidth}
        \centering
        \begin{tabular}{lcc}
        \toprule
        Code Preprocessing & Held-out rubric & Held-out exam \\
        \midrule
        None & 70.9 & 63.3 \\
        Obfuscation & 73.3 & 66.9 \\
        Byte-pair & \textbf{74.2} & \textbf{69.8} \\
        \bottomrule
        \end{tabular}
        \caption{\textbf{Ablation:} code preprocessing.}
    \end{subtable}
    \begin{subtable}{.45\linewidth}
        \centering
        \begin{tabular}{lcc}
        \toprule
        & \multicolumn{2}{c}{AP}  \\
        Aug \% & Held-out rubric & Held-out exam \\
        \midrule
        1\% & 72.2 & 68.1 \\
        5\% & 76.4 & 69.6 \\
        10\% & \textbf{78.2} & \textbf{72.6} \\
        25\% & 77.7 & 68.2\\
        50\% & 75.0 & 65.3 \\
        100\% & 70.0 & 64.8 \\
        \bottomrule
        \end{tabular}
        \caption{\textbf{Ablation:} task augmentation amount.}
        \label{table:app:ablation1}
    \end{subtable}
    \begin{subtable}{.45\linewidth}
        \centering
        \begin{tabular}{lcc}
        \toprule
        & \multicolumn{2}{c}{AP}  \\
        \# shots & Held-out rubric & Held-out exam \\
        \midrule
        1-shot & 57.8 & 58.9 \\
        2-shot & 64.5 & 60.1 \\
        5-shot & 76.1 & 66.9 \\
        10-shot & \textbf{80.5} & \textbf{71.4} \\
        \bottomrule
        \end{tabular}
        \caption{\textbf{Ablation:} number of shots.}
        \label{table:app:ablation2}
    \end{subtable}
    \caption{Albation experiments testing the impact of different components of the ProtoTransformer on model performance.}
\end{table}

\section{Analysis of Embeddings}
\label{sec:embedding}
\begin{figure*}[h!]
\centering
\begin{subfigure}[b]{0.24\linewidth}
    \centering
    \includegraphics[width=0.5\linewidth]{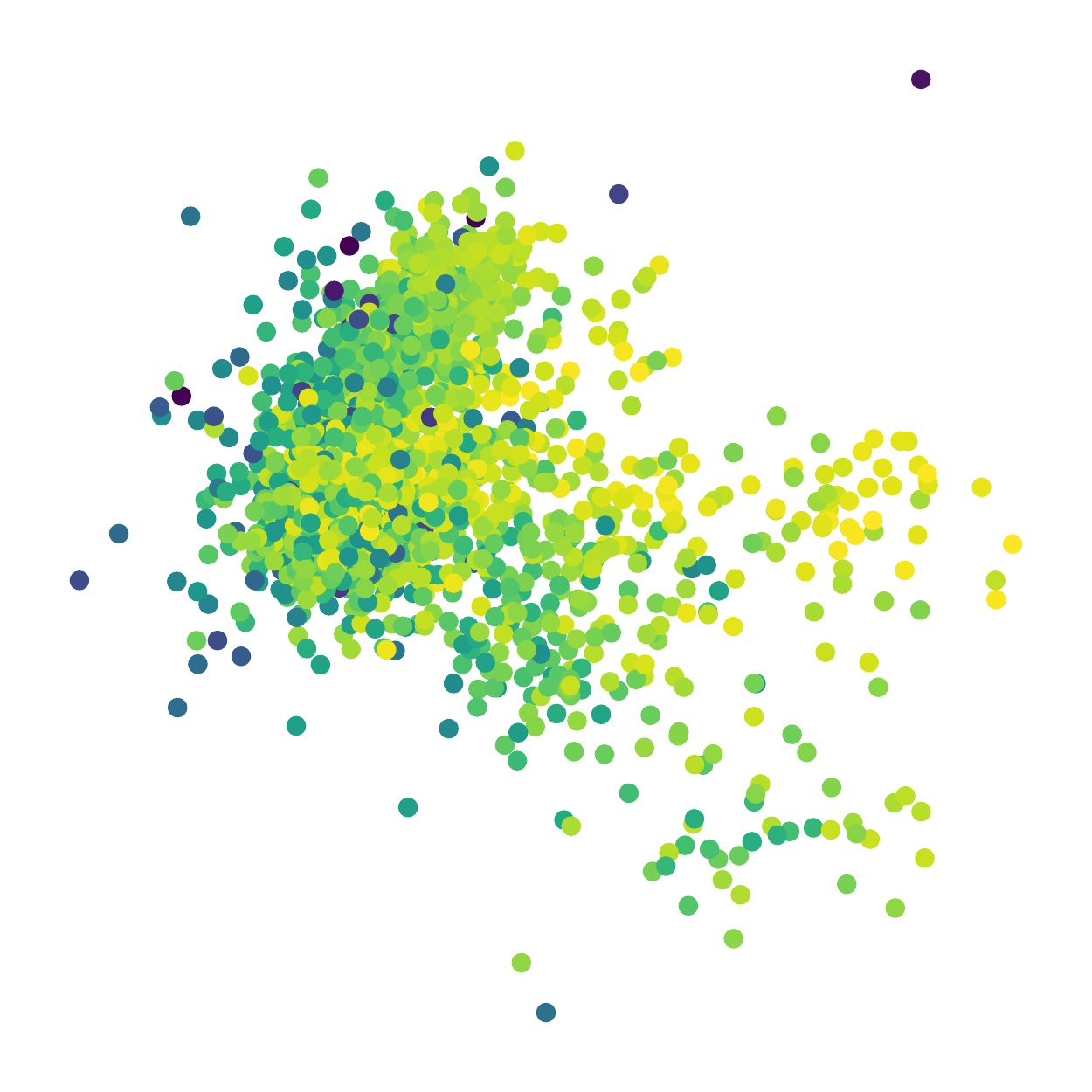}
    \caption{All Students}
\end{subfigure}
\begin{subfigure}[b]{0.24\linewidth}
    \centering
    \includegraphics[width=0.5\linewidth]{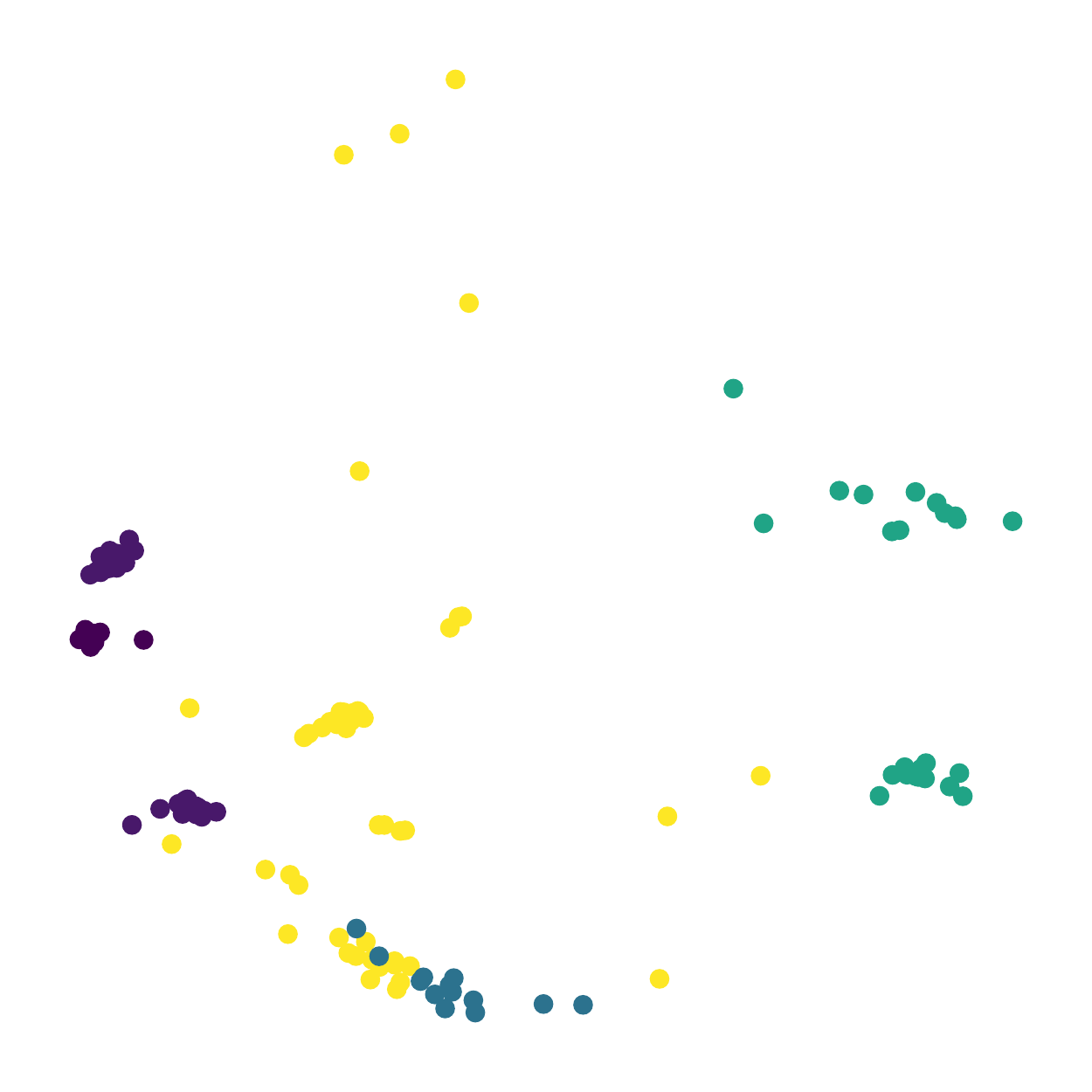}
    \caption{Random Student}
\end{subfigure}
\begin{subfigure}[b]{0.24\linewidth}
    \centering
    \includegraphics[width=0.5\linewidth]{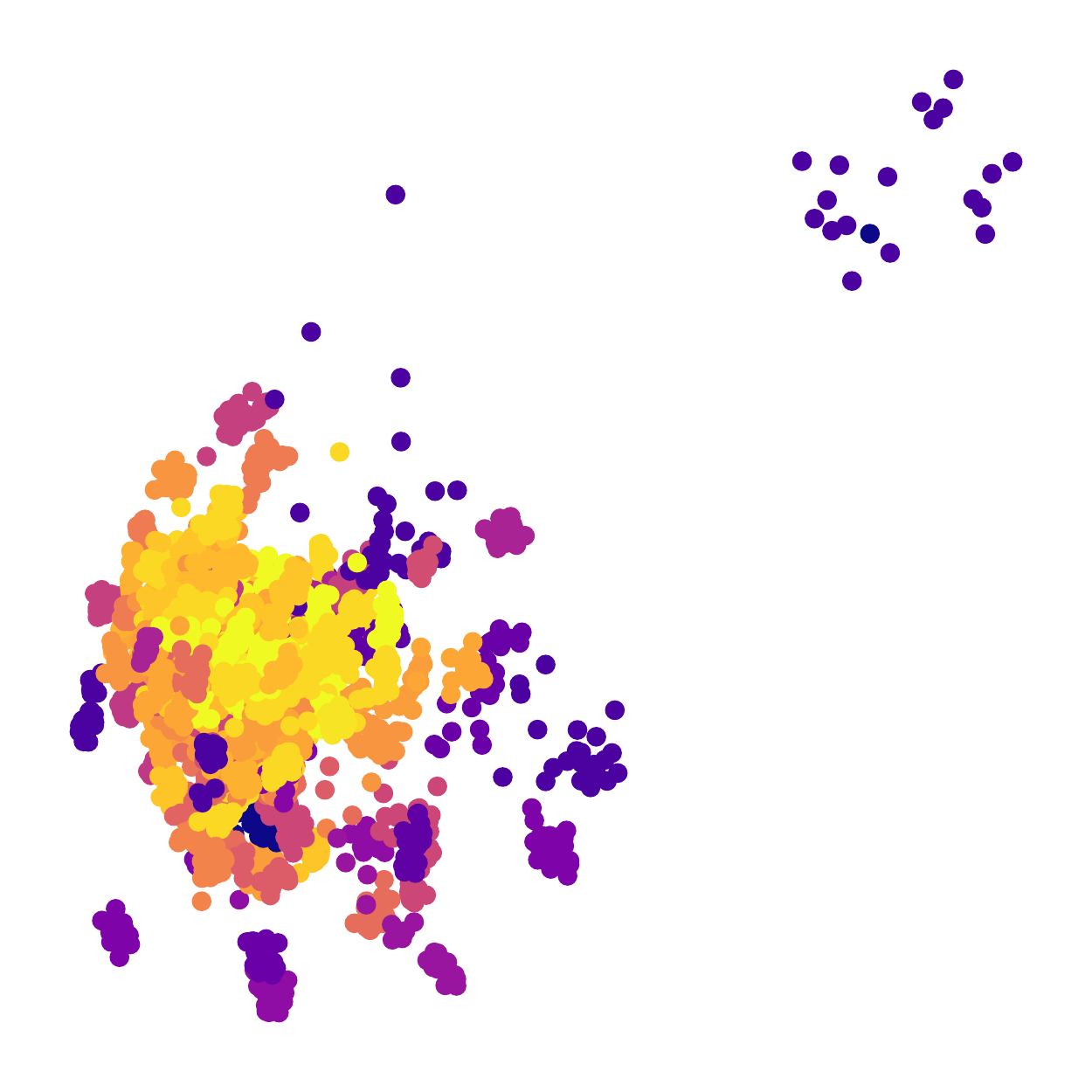}
    \caption{Random Question}
\end{subfigure}
\begin{subfigure}[b]{0.24\linewidth}
    \centering
    \includegraphics[width=0.5\linewidth]{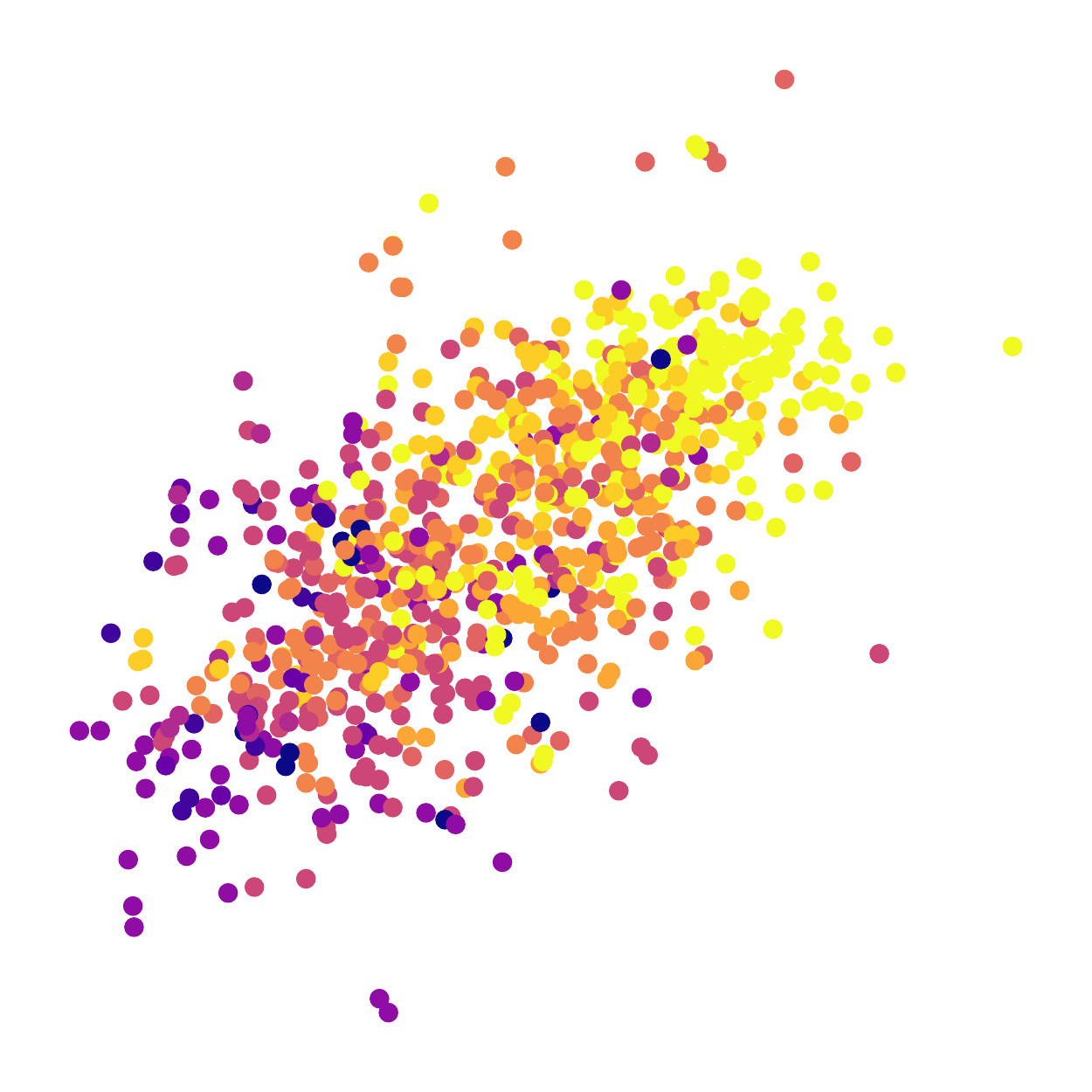}
    \caption{Random Question}
\end{subfigure}
\caption{{\small \textbf{Embedding clustering:} The color represents the true grade (darker is lower).
In (a), we visualize the average embedding (over questions) for every student.
 Subfigure (b) shows embeddings for each question for a random student. Subfigures (c) and (d) show all student solutions for two randomly chosen questions.}}
\label{fig:cluster}
\end{figure*}

We can explore what the ProtoTransformer has learned about student code by exploring clusters in the shared embedding space.
Figure~\ref{fig:cluster} shows a PCA projection of embedding solutions to two dimensions.
For each solution, we are given a true numeric grade assigned by a teaching assistant (between 0 and 100), which is not used in training nor evaluation.

For every student, we first compute the average  embedding over all questions, which we can interpret as a measure of student understanding. We see that Figure~\ref{fig:cluster}a uncovers a linear relationship as students in the top-left corner have lower performance.
In Figure~\ref{fig:cluster}b we focus on a single student, visualizing their embedded solutions. We observe distinct clusters that separate the questions that this student excels at versus those this student does poorly on.
This information could be useful for an instructor in measuring learning and growth.

Similarly, we visualize all student attempts for a single question in Figures~\ref{fig:cluster}c \& d, finding surprisingly different structure across questions.
Figure~\ref{fig:cluster}c contains a large cluster of high-scoring solutions with several distinct clusters of low-scoring solutions.
In contrast, Figure~\ref{fig:cluster}d contains a single cluster with solutions scoring higher from bottom-left to top-right in a continuous manner.
In practice, this analysis could help instructors evaluation question difficulty and curriculum design.

\section{NLP Experiment Details}
\label{sec:supp:nlp}

In all cases, we fit the ProtoTransformer with Adam for 30 epochs with weight decay of 0.01, $\beta_1$=0.9, $\beta_2$=0.999, and a linear learning rate schedule to 1e-4 with 10k steps warmup \citep{liu2019roberta}.
The LSTM baseline was bi-directional with 4 stacked layers. Pretraining weights were taken a RoBERTa model from HuggingFace.

\subsection{Additional NLP Experiments}

The results in the main paper split meta-training and meta-test tasks by topic such that tasks in the meta-test set use topics not seen in meta-training.
Here we run a similar study in which we ensure that no combination of classes appearing in a meta-training task appear in a meta-test task.
That is, the meta-test set might contain all 20 topic classes but every meta-test task of $N$ classes is novel e.g. \texttt{comp.graphics}, \texttt{sci.crypt}, and \texttt{alt.atheism} form a 3-way task that is unique if this exact combination does not appear in the meta-training set, although each class individually may be used in tasks for meta-training.
Note that this is an easier setup than the experiment in the main paper.
Examples for meta-training tasks are taken from training split of 20-newsgroups whereas examples for meta-test tasks are taken from the test split.
All hyperparameters and architecture choices are as in supplement ~\ref{sec:supp:nlp}.

\begin{table}[h!]
    \small
    \centering
    \begin{tabular}{lcccc}
    \toprule
    & \multicolumn{2}{c}{2-way} & \multicolumn{2}{c}{5-way} \\
    \midrule
    Model & 1-shot & 2-shot & 1-shot & 2-shot \\
    \midrule
    All & \textbf{92.4} & \textbf{93.2} & 82.2 & \textbf{85.2} \\
    No Pretrain & 62.1 & 80.1 & 55.7 & 63.5 \\
    No Side & 89.5 & 91.5 & 76.8 & 82.2\\
    No Cloze & 88.4 & 90.9 & 80.3 & 83.3\\
    LSTM & 62.5 & 64.2 & 38.1 & 43.6 \\
    Matching & -- & 92.2 & -- & 82.3 \\
    \bottomrule
    \end{tabular}
    \caption{\textbf{Few-shot topic classification:} Performance of the ProtoTransformer Network with various ablations on predicting sentence topic using the 20-newsgroups dataset.}
    \label{table:supp:nlp}
\end{table}

\section{Runtime and Cost}
\label{sec:app:runtime}
Training the ProtoTransformer for 300 epochs on the dataset of university coursework takes 12 to 13 hours on a single Titan X GPU. Finetuning 3 layers of a pretrained RoBERTa model fits within the 11Gb memory.
All models, baselines, and ablations used the same compute resources. A supervised baseline for a single task takes roughly 15 to 30 minutes to train.

\section{Training Tools}
\label{sec:tools}
We use the Huggingface \texttt{transformers} library \citep{wolf-etal-2020-transformers} for a RoBERTa implementation, which is under a Apache 2.0 license. We use the CodeBERT repository (\url{https://github.com/microsoft/CodeBERT}), which is under a MIT license. For our infrastructure, we use PyTorch and PyTorch Lightning \citep{falcon2019pytorch}, which have BSD and Apache 2.0 licenses, respectively.